\documentclass{article}
\usepackage{amsfonts}

\usepackage{graphics}

\newtheorem{theorem}{Theorem}
\newtheorem{definition}{Definition}
\newtheorem{lemma}{Lemma}
\newtheorem{proposition}{Proposition}
\newtheorem{remark}{Remark}

\input{tcilatex}
\begin{document}

\title{Gibbs and Quantum Discrete Spaces}
\author{V. A. Malyshev}
\date{}
\maketitle

\begin{abstract}
Gibbs field is one of the central objects of the modern probability,
mathematical statistical physics and euclidean quantum field theory. Here we
define and study its natural generalization for the case when the space,
where the random field is defined is itself random. Moreover, this
randomness is not given apriori and independently of the configuration, but
rather they depend on each other, and both are given by Gibbs procedure; We
call the resulting object a Gibbs family because it parametrizes Gibbs
fields on different graphs in the support of the distribution. We study also
quantum (KMS) analog\ of Gibbs families.
\end{abstract}

\tableofcontents

\section{Introduction}

This paper had different motivations.

Gibbs field \cite{dob, laru} is one of the central objects of the modern
probability theory, mathematical statistical physics and euclidean field
theory \cite{glja}. Here we define its natural generalization, when the
background space (lattice, graph), where the field is defined, is itself
random. Moreover, this randomness is not given apriori and independently of
the configuration, but the space and the configuration on it depend on each
other and are given by a kind of Gibbs construction. The resulting
distribution is called a Gibbs family because it parametrizes finite of
continuum set of pairs $(G,\mu )$, where $G$ is a graph, $\mu $ is some
Gibbs measure on $G$, belonging to the support of the distribution. Even if
the set of configurations is trivial, one gets a nontrivial object, which is
called a Gibbs graph.

Quantum analog of Gibbs families is a generalization of quantum spin system
to the case of ''quantum'' lattice \cite{brro}. If there is no spin we also
get an interesting object called quantum graph (lattice,complex, space).

We give here mathematical foundations of such theory, which includes many
different aspects - from the general notion of locality to combinatorics of
graphs.

Another motivation were questions which appear naturally while thinking
about analytic and probabilistic motivations of such modern physical
theories as string theory, $M$-theory, and many other approaches to quantum
gravity. These theories now are on the physical level of rigorousness,
having however many deep algebraic and geometric mathematics.

First obvious question is ''what is locality if there is no background space
?'' The space in physical theories normally is given apriori - as the scene
for all further actions. For a long time only three dimensional euclidean
space were used in physics. In the 20 century the dimension $d$ was allowed
to vary, also arbitrary smooth manifolds were studied instead of $R^{d}$.
Moreover, the global structure of the manifold is can may depend on the
matter fields via the general relativity equations, restrictions of the
string theory etc. However, the postulate of the local smoothness was
unquestionable, and rests as it is until now in many modern theories. This
postulate, being very convenient for differential geometry techniques, is
the main source of divergences in quantum field theory and difficulties with
attempts to quantize the metrics itself. The logics of things puts natural
questions of what could be a replacement for local euclidean postulate.

At present many working in the field are looking answers in notions of
discrete and noncommutative \cite{connes, landi, lali, manin} spaces.
However, even in physics discrete spaces are considered to be subsets of
some ''background space'', and noncommutative spaces are smooth deformations
of locally euclidean spaces. Moreover, there is a tendency in physics to put
the space on equal footing with matter fields. In particular, this means
that it should be random or quantum.

It appears immediately that for general discrete spaces the mere notion of
locality can have different definitions, which are discussed here. Locality
in physics is assiated with the Markov-Gibbs property. The ''local'' space
is constructed from discrete elements (quanta) via a special (Gibbs)
procedure. There are two main possibilitires to do this. If one assumes that
a neighborhood of the point may have any number of quanta then to get
classical spaces one needs renormalization procedures similar to
renormalization in quantum fields theory. If one understands locality in
more restrictive ''metric'' sense, that is when a quantum is a neighborhood
itself, then one does not need renormalizations. The latter understanding is
not quite satisfactory because the metrics itself may vary sufficiently.

The plan of the paper is the following. In section 2.1 we introduce two
definitions of finite Gibbs discrete spaces (Gibbs families). First of all
we introduce a simple and useful notion of a graph with a local structure,
it includes simplicial complexes and many other general ''space-like''
objects. The first definition (local) \ is when the energy of the graph is
equal to the sum of energies of its subgraphs, the second one (superlocal),
when the energy is the sum of energies of all neighborhoods of a given
radius. It is shown that these two definitions can be reduced to each other
by only using a procedure, similar to renormalization in the quantum field
theory.

In section 2.2 we consider limiting Gibbs families, give the analog of the
Dobrushin-Lanford-Ruelle condition for Gibbs fields and the compactness
condition. For a given Gibbs family the conditional measure (for a fixed
graph $G$) on the configurations on $G$ is a Gibbs measure in the standard
sense with the same potential.

In section 2.3 we discuss an important subtle question ''what is a random
countable graph ?''. One of the possibilities is to assume the existence of
a fixed vertex (origin). If not, then one could (very unnaturally) assume a
common enumeration of the vertices af each countable graph, also to consider
countable graphs imbedded to a common ''background'' space. However, without
such assumptions the Kolmogorov methods to define a probability measure via
finite-dimensional distributions does not work. We have thus to define a
substitute for probability measures on the equivalence classes of countable
graphs. This analog is called an empirical distribution, and we give
examples of such distributions from discrete quantum gravity.

In section 3 we discuss the asymptotics of partition functions for different
models of discrete spaces. In particular we get the logarithmic asymptotics
of $\sum_{g}F_{b,p}(g)$ for $p\rightarrow \infty ,b=\alpha p$, where $%
F_{b,p}(g)$\ is the number of maps with $p+1$ vertices and $b+p$ edges on
the closed compact surface of genus $g$. Note that a particular case of a
map is a triangulation, and the experience for censoring maps on a surface
with fixed genus shows similar asymptotical behaviour for different classes
of maps.

In section 4 we consider phase transitions.\ In section 4.2 we study in
detail the simplest model of pure planar gravity (with one boundary) where
one has a complete control: existence of the thermodynamic limit, explicit
free energy, complete phase diagram, including the critical point. We have
here also an effect of the influence of the observer on fluctuations of the
boundary in the critical point. The corresponding model with two boundaries
is studied only in the ''high temperature'' case. In section 4.3 we give an
example of the topological phase transition in a locally homogeneous graph,
where all vertices have isomorphic 1-neighborhoods. In 4.4 we prove the
uniqueness of the limiting Gibbs family for a model with factorial growth of
the partition function. This allows to give a local Gibbs characterization
of tress and free noncommutative groups. In 4.5 we discuss how different
topologies could appear on different scales. The notion of scale includes
thermodynamicn hydrodynamic limits etc.

In section 5.1 we introduce quantum analogs of Gibbs graphs with a local
structure - quantum grammars on graphs. We prove basic results concerning
selfadjointness of the hamiltonian, existence of the automorphism group of
the corresponding $C^{\ast }$-algebras. Quantum spin systems are a very
particular case of this construction. For linear graphs \ in the high
temperature region we prove existence and uniqueness of KMS states. In 5.2
we give many examples of $C^{\ast }$-algebras and hamiltonians for quantum
grammars: linear grammars and Toeplitz operators, quantum expansion and
contraction of the space, Lorentzian models etc.

\section{Gibbs Families}

\subsection{Finite Gibbs families}

\subparagraph{Spin graphs}

We consider graphs $G$ (finite or countable) with the set $V=V(G)$ of
vertices and the set $L=L(G)$\ of edges. It is always assumed that there is
at most one edge between any two vertices and that each vertex has a finite
degree (the number of incident edges). Graphs are assumed to be connected,
no loops, unless otherwise stated. However, general results can be easily
reformulated for graphs with loops and multiple edges. In this case the loop
adds number $2$ to the degree of the vertex.

Subgraph of the graph $G$ is defined by a set of vertices $V_{1}\subset V$
together with some edges, inherited from $L$ and connecting some vertices
from $V_{1}$. A subgraph $G(V_{1})$ of the graph $G$ with the set $V_{1}$ of
vertices is called regular if it contains each edge in $L(G)$ from the
vertices of $V_{1}$.

The set $V$ of vertices is a metric space, if the distance $d(x,y)$ between
vertices $x,y\in V$ is defined as the minimal length (number of vertices) in
a path, connecting these vertices.

Spin graph $\alpha =(G,s)$ is defined as some graph $G$ together with a
function $s:V\rightarrow S$, where $S$ is some set (spinvalues). Graphs and
spin graphs are always considered up to an isomorphism that is no
enumeration of vertices is assumed.\ Isomorphism is a one-to-one
correspondence between vertices, respecting edges and spins. Let $\mathcal{G}%
_{N}$ ($\mathcal{A}_{N}$) be the set of equivalence classes (with respect to
isomorphisms) of connected graphs (spin graphs) with $N$ vertices. Countable
graphs correspond to $N=\infty $.

Further on we use a somewhat free terminology: we say graph or spin graph
instead of equivalence class, subgraph instead of spin subgraph (thus
omitting the adjective spin) etc.

\subparagraph{Spin graphs with the origin}

We shall also consider graphs with a specified vertex (rooted graphs),
called the origin or the root. In this case all isomorphisms are assumed to
respect the origin. When one has to stress that the considered graphs have
the root $v$, we shall write $G^{(v)},\mathcal{G}_{N}^{(v)}$ etc.

Define the annular neighbourhood $\gamma (\alpha ,v;a,b)$ of vertex $v$ to
be the regular subgraph of the spin graph $\alpha $ with the set of vertices 
$V(\gamma (\alpha ,v;a,b))=\{v^{\prime }:a\leq d(v,v^{\prime })\leq b\}$. $%
O_{d}(v)=\gamma (\alpha ,v;0,d)$ is called the $d$-neighborhood of $%
v,O(v)=O_{1}(v)$. $R_{0}(G)=\max_{v\in V(G)}d(0,v)$ is the radius of $G$
with respect to vertex $0$. In particular, $R_{v}(O(v))=1$. Then the
diameter of the graph $G$ is $d(G)=\max_{v,v^{\prime }\in V(G)}d(v,v^{\prime
})=\max_{v\in V(G)}R_{v}(G)$.

Denote $\mathcal{G}_{N}^{(0)}$ ($\mathcal{A}_{N}^{(0)}$) the set of
equivalence classes of connected graphs (spin graphs) with root $0$ and
radius $N$ with respect to $0$.

Denote $\mathcal{A}^{(0)}=\cup _{N\neq \infty }\mathcal{A}_{N}^{(0)}$ the
set of all finite graphs with root $0$.

\subparagraph{Graphs with a local structure}

Spin graph is a particular case of a graph with a local structure. This
notion is sufficient for description of any multidimensional objects.

\begin{definition}
Let the function $s(\gamma )$ on $\mathcal{A}_{d}^{(0)}$ be given together
with the function $t(\gamma )$\ on the set $\mathcal{A}(d)$\ of spin graphs
of diameter $d$\ with values in some set $S$. Superlocal structure (of
radius $d$) on $G$ is given by the set of values $s(O_{d}(v)),v\in V(G)$.
Local structure (of diameter $d$) on $G$ is given by the set of values $%
t(\gamma )$ for all regular graphs $\gamma $ of diameter $d$.
\end{definition}

Examples:

\begin{itemize}
\item  Spin graphs correspond to a local (or superlocal) structure with $d=0$%
. Many definitions and results for spin graphs, obtained below, can be
generalized on the case $d>0$ by simple rewriring.

\item  Gauge fields on graphs: for each vertex and for each edge values from
a group $R$ are defined, in this case one can take $d=1$.

\item  Simplicial complex is completely defined\textit{\ by the function on
its one-dimensional skeleton}: it takes value $1$ on each complete regular
subgraph $\gamma $ of diameter $1$, iff this subgraph defines a simplex of
the corresponding dimension, and $0$ otherwise. Here one can also take $d=1$.

\item  Penrose quantum networks \cite{penr1, baez1}: in a finite graph, each
vertex of which has degree $3$, to each edge $l$ some integer $p_{l}=2s_{l}$
is prescribed, where halfintegers $s_{l}$\ are interpreted as degrees of
irreducible representations of $SU(2)$. Moreover, in each vertex the
following condition is assumed satisfied: the sum of $p_{i}$, for all three
incident edges, is even and any $p_{i}$ does not exceed the sum of two other
values. Then for the tensor product of three representations there exists a
unique (up to a factor) invariant element, which is prescribed to this
vertex.

\item  However, there are structures on graphs which cannot be characterized
as graphs with a local structure. Examples are trees, Cayley graphs etc. We
shall see later, however, that Gibbs characterization (also local in some
sense) is possible.
\end{itemize}

\subparagraph{$\protect\sigma $-algebra and free measure}

Let $\mathcal{A}$ be an arbitrary set of finite spin graphs $\alpha =(G,s)$.
Let $\mathcal{G}=\mathcal{G}(\mathcal{A})$ be the set of all graphs $G$ for
which there exists a function $s$ with $\alpha =(G,s)\in \mathcal{A}$. It is
always assumed that if $\alpha =(G,s)\in \mathcal{A}$, then for any $%
s^{\prime }$ all $(G,s^{\prime })$ also belong to $\mathcal{A}$.

Then $\mathcal{A}$ is a topological space, a discrete (finite or countable)
union of topological spaces $T_{G}=S^{V(G)},G\in \mathcal{G}(\mathcal{A})$.
Borel $\sigma $-algebra on $\mathcal{A}$ is generated by the cylindrical
subsets $A(G,B_{v},v\in V(G)),$ $G\in \mathcal{G}$, where $B_{v}$ are some
Borel subsets of $S$. Moreover, $A(G,B_{v},v\in V(G))$ is the set of all $%
\alpha =(G,s)$ such that for some fixed graph $G$ the functions
(configurations) $s(v):V(G)\rightarrow S$ satisfy the following condition: $%
s(v)\in B_{v}$ for all $v\in V(G)$.

Let on $S$ a nonnegative (not necessary probabilistic) measure $\lambda _{0}$
be given. The free measure is the following nonnegative measure $\lambda _{%
\mathcal{A}}$ on $\mathcal{A}$, defined by 
\begin{equation}
\lambda _{\mathcal{A}}(A(G,B_{v},v\in V(G)))=\prod_{v\in V(G)}\lambda
_{0}(B_{v})
\end{equation}

\subparagraph{Potentials}

Potential (local) is a function $\Phi :\cup _{N<\infty }\mathcal{A}%
_{N}\rightarrow R\cup \left\{ +\infty \right\} $, that is a function on the
set of finite spin graphs, invariant with respect to isomorphisms. If $\Phi $
can take value $\infty $, then we say that $\Phi $ has a hard core.

We say that $\Phi $ has a finite diameter, if $\Phi (\alpha )=0$ for all $%
\alpha $ with diameter greater than some $d$. Minimal such $d$ is called the
diameter of $\Phi $.

For the potential $\Phi $ the energy of the subgraph $\alpha $ is defined as 
\begin{equation}
H(\alpha )=\sum_{\gamma \subset \alpha }\Phi (\gamma ),\alpha =(G,s)
\end{equation}
where the sum is over all regular connected subgraphs $\gamma $ of the spin
graph $\alpha $.

A particular case is the chemical potential. It is given by the function $%
\Phi $, equal to some constant $\mu _{0}$ for each vertex $v$, indepedently
of the spin in $v$, and equal to some constant $\mu _{1}$ for each edge, and
equal $0$ in other cases. Otherwise speaking the chemical potential is 
\begin{equation}
H_{N}(\alpha )=\mu _{0}V(\alpha )+\mu _{1}L(\alpha )
\end{equation}
where $V(\alpha ),L(\alpha )$ are the numbers of vertices and edges of $%
\alpha $.

\subparagraph{Finite Gibbs families}

Let $\mathcal{A}$\ be some set of finite spin graphs. Gibbs $\mathcal{A}$%
-family with potential $\Phi $ is a probability measure $\mu _{\mathcal{A}}$
on $\mathcal{A}$, defined by the following density withy respect to the free
measure 
\begin{equation}
\frac{d\mu _{\mathcal{A}}}{d\lambda _{\mathcal{A}}}(\alpha )=Z^{-1}\exp
(-\beta H(\alpha )),\alpha \in \mathcal{A}
\end{equation}
where $\beta \geq 0$ is the inverse temperature. Moreover it is assumed that 
$Z\neq 0$, that is $H(\alpha )<\infty $ at least for one $\alpha \in 
\mathcal{A}$. Below we give examples when $Z=0$. We shall assume also that,
when $\mathcal{A}$ is infinite, the following stability condition holds 
\begin{equation}
Z=Z(\mathcal{A})=\int_{\mathcal{A}}\exp (-\beta H(\alpha ))d\lambda _{%
\mathcal{A}}<\infty
\end{equation}

If $S$ is trivial (consists of one element), we will get a probability
distribution on the set of graphs, which will be called a Gibbs graph.

\subparagraph{Stability}

Potential $\Phi $ is said to be stable on $\mathcal{A}_{N}^{(0)}$ if for any 
$N$ the partition function $Z_{N}<\infty $. Potential $\Phi \equiv 0$ is
obviously not stable, and the chemical potential with $\mu _{0}=0$ as well.

\begin{lemma}
The chemical potential $\mathcal{A}_{N}^{(0)}$ is not stable for any finite $%
\mu _{0},\mu _{1}$.
\end{lemma}

Proof. Let $g(n,k)$ be the number of graphs of radius $1$, which have the
origin $0$ of degree $n$, and moreover, the number of edges between other
vertices ia $k$. It is sufficient to prove that 
\[
G_{n}=\sum_{k=0}^{N}g(n,k)\exp (-\mu _{1}(k+n)-\mu _{0}(n+1)) 
\]
grows together with $n$, where $N=\frac{n(n-1)}{2}$ is the number of
different vertices of $n$ enumerated vertices.

In fact, the number of graphs with $n$ enumerated vertices and $k$ edges
equals $C_{N}^{k}$ . At the same time the number of ways to enumerate $n$
vertices is not greater than $n!=2^{n\log n+O(n)}$. Let us prove that for
any $\mu _{1}$ there exists sufficiently small $\delta >0$ such that for $%
k=\delta N,n\rightarrow \infty $ \ one has $G_{n}\rightarrow \infty $. We
have 
\[
g(n,k)\exp (-\mu _{1}(k+n)-\mu _{0}(n+1))\approx 
\]
\[
\approx 2^{-N(\delta \ln \delta +(1-\delta )\ln (1-\delta ))+O(n\log n)}\exp
(-\mu _{1}(k+n)-\mu _{0}(n+1))> 
\]
\[
>\exp (-\mu _{1}\delta N)2^{N(-\delta \ln \delta +\delta )+O(n\log
n)}\rightarrow \infty 
\]
if $\delta $ is sufficiently small.

Instability can appear because of big degrees of vertices. That is why
regularization is necessary. One of the possibilities is to introduce
regularizing potentials.

An example of regularizing potential is the following potential. Let $\gamma
_{2}$ ($\gamma _{3}$) be the graph consisting of three vertices $1,2,3$ and
two edges $12,13$ (three edges $12,13,23$). Potential $\Phi _{2}=\mu
_{2}P(\alpha )$, where $P(\alpha )$ is the number of regular subgraphs of $%
\alpha $, isomorphic to $\gamma _{2}$ or to $\gamma _{3}$, garanties the
stability of the potential $\Phi +\Phi _{2}$ for many $\Phi $. For exmple,
the following proposition holds, which is not difficult to prove.

\begin{proposition}
If $\mu _{2}>\ln 2$, then the potential $\Phi =\Phi _{0}+\Phi _{1}+\Phi _{2}$
is stable for any $\mu _{0},\mu _{1}\geq 0$.
\end{proposition}

However', more interesting potentials are those which are equal $\infty $ on
graphs of special kind. For example, it is infinite on all graphs with
diameter greater or equal $1$ or $2$, which also have at least one vertex of
degree $r+1$. Such potential is equivalent to the apriori restricting
ourselves to the class of graphs, eahc vertex of which has degree not
exceeding $r$.

Another possibility is to revise the notion of locality.

\subparagraph{Locality and Superlocality}

In many papers on discrete quantum gravity it is assumed apriori that the
action takes place on a smooth or piecewise linear manifold of some fixed
dimension. But then locally the space is classical. More consistent program
could be to refuse from such assumptions. But then the central question
arises: what is locality if there is no background space?

One sees immediately that locality can be understood in many ways. Firstly,
local can be an object, if it is defined in terms of ''small'' connected
subgraphs. An essentially stronger notion of locality - superlocality -
arises when we know that such subgraph coincides with some neighbourhood of
a vertex. Accordingly, call a superlocal potential of radius $r$ a function $%
\Phi :\mathcal{A}_{r}^{(0)}\rightarrow R\cup \left\{ +\infty \right\} $,
that is a function on the set of finite rooted spin graphs of radius $r$
with respect to $0$, invariant with respect to the isomorphisms.

The energy of the superlocal potential is defined as 
\[
H(\alpha )=\sum_{v\in V(\alpha )}\Phi (O_{r}(v)), 
\]
where the sum is over all vertices of the spin graph $\alpha $. In this case
we shall call Gibbs families superlocal, and the Gibbs families defined
above will be called local.

Superlocal potential can be reduced to a local one, but only with some
additional limiting procedure, see below. One can say that if one wants to
describe the world locally but without apriori restrictions on the structure
of the graphs, then we should use renormalizations.

To construct examples the following classes of graphs are useful. Let us
consider a set of ''small'' graphs $G_{1},...,G_{k}$ with root $0$ and
having radius $r$ with respect to $0.$ We shall say that graph $G$ is
generated by graphs $G_{1},...,G_{k}$, if any vertex of $G$ has a
neighborhood (of radius $r$) isomorphic to some of the generators $%
G_{1},...,G_{k}$. Denote $\mathcal{G}(G_{1},...,G_{k})$ the set of all such
graphs. Introduce a superlocal potential $\Phi =\Phi _{G_{1},...,G_{k}}$ of
radius $r$, putting $\Phi (\Gamma )=0$, if the rooted subgraph $\Gamma $ is
isomorphic to one of the $G_{i}$, and $\Phi (\Gamma )=\infty $ otherwise. It
is obvious that any Gibbs family with potential $\Phi $ has the support on $%
\mathcal{G}(G_{1},...,G_{k})$.

\subparagraph{Example of $Z_{N}=0$}

Even finite Gibbs families with potential $\Phi =\Phi _{G_{1},...,G_{k}}$ do
not necessarily exist. As the simplest example consider the case with $%
k=1,r=1$, when $G_{1}$ is a finite complete graph $g_{m}$ of radius $1$ with 
$m+1$ vertices $0,1,2,3,...,m$ and root $0$. Then it is ''selfgenerating''
and $\mathcal{G}(G_{1})$ consists of one element - $g_{m}$ itself.

Another example. Introduce the graphs $g_{m,0}$ with $m+1$ vertices $%
0,1,2,3,...,m$ and $m$ edges $01,02,...,0m,$ and $g_{m,m}$ with $m+1$
vertices $0,1,2,3,...,m$ and $2k$ edges $01,02,...,0m,12,23,...,m1$. One can
see that the pentagon $G_{1}=g_{5,5}$ cannot generate a countable graph.

\subparagraph{Renormalization}

Superlocal potential can be expressed with local potentials in the following
way. We shall explain this for the case $k=1$, let the radius of $G_{1}$ be $%
r$. We shall construct a finite Gibbs family with support on $\mathcal{G}%
(G_{1})$, without using superlocal potentials. Let $F(G_{1})$ be the set of
all graphs, isomorphic to some proper regular subgraph $\gamma $ of $G_{1}$
and such that the radius of $\gamma $ with respect to some vertex $v\in
V(\gamma )$ equals $r$. Let $F(\Gamma ,\gamma )$ be the set of all proper
regular subgraphs of $\Gamma $, isomorphic to $\gamma $.

Introduce the ''cut-off'' local potential $\Phi _{R}$ as follows:

\begin{itemize}
\item  $\Phi _{R}(\gamma )=\infty $ for graphs $\gamma $, having radius $r$
with respect to some of its vertices and such that $\gamma \neq G_{1}$ and $%
\gamma \notin F(G_{1})$;

\item  $\Phi _{R}(\gamma )=0$ for graphs which do not have radius $r$ for
any of its vertices;

\item  $\Phi _{R}(\gamma )=R,R>0,$ for any graph $\gamma \in F(G_{1})$,

\item  $\Phi _{R}(G_{1})=0$.
\end{itemize}

The limit for $R\rightarrow \infty $\ of the Gibbs family $\mu _{\mathcal{A}%
}(R)$ with potential $\Phi _{R}$ may not exist, but the following theorem
holds.

\begin{theorem}
Assume that $\mathcal{G}(G_{1})\cap \mathcal{A}_{N}\neq \emptyset $. Then
there is such $k$\ and such local potentials $\phi _{R}^{(j)},j=1,2,...k$,
that the limit as $R\rightarrow \infty $ of the Gibbs family on $\mathcal{A}%
_{N}$\ with renormalized potential 
\[
\Phi _{R}^{(ren)}=\Phi _{R}+\sum_{j=1}^{k}\phi _{R}^{(j)} 
\]
exists and has its support on $\mathcal{G}(G_{1})\cap \mathcal{A}_{N}.$
\end{theorem}

Proof. Define the potentials (counterterms) $\phi _{R}^{(j)}$, equal zero
everywhere except for the following cases:

\begin{itemize}
\item  if $\Gamma $ is isomorphic to $G_{1}$, then 
\[
\phi _{R}^{(1)}(\Gamma )=\sum_{\gamma \in F(G_{1})}\phi _{R,\gamma
}^{(1)}(\Gamma ),\phi _{R,\gamma }^{(1)}(\Gamma )=-R\sum_{\Gamma \in
F(G_{1},\gamma )}1 
\]

\item  for any $j\geq 2$, if $\Gamma $ is the union of $j$ pairwise
different subgraphs $\Gamma _{i},i=1,...,j,$ where each $\Gamma _{i}$ is
isomorphic $G_{1}$, then 
\[
\phi _{R}^{(j)}(\Gamma )=\sum_{\gamma \in F(G_{1})}\phi _{R,\gamma
}^{(j)}(\Gamma ),\phi _{R,\gamma }^{(j)}(\Gamma )=(-1)^{j}R\sum_{\Gamma \in
F(\Gamma _{1},\gamma )\cap ...\cap F(\Gamma _{j},\gamma )}1 
\]
\end{itemize}

Note that it follows from the definition that $\phi _{R}^{(j)}=0$ starting
from some $j$. Consider a vertex $v$. If its neighborhood $O_{r}(v)$ does
not belong to $F(G_{1})$, then $\Phi _{R}$ equals $\infty $, and thus it is
impossible. Let now the neighborhood $O_{r}(v)$ is not isomorphic to $G_{1}$%
\ and thus belongs to $F(G_{1})$. Then the potential $\Phi _{R}$ is $R$ on
it, and the counterterms equal $0$. It follows that $\Phi _{R}^{(ren)}$
equals $R$ as well. Consider now some subgraph $\gamma \in F(G_{1})$, which
does not belong to exactly $m$ subgraphs $\Gamma _{i}$, isomorphic to $G_{1}$%
. Then by inclusion-exclusion formula, for any $\gamma $ 
\[
\Phi _{R}(\gamma )+\sum_{j=1}^{m}\phi _{R,\gamma }^{(j)}(g)=0 
\]
Thius the energy of the graphs from $\mathcal{G}(G_{1})\cap \mathcal{A}_{N}$
equals $0$, and for the remaining graphs it is equal or greater than $R$,
this give the result.

Note that $\phi _{R}^{(j)}(\Gamma )$\ are similar to counterterms of the
standard renormalization theory.\ For $k>1$ more complicated combinatoris
appears. It could be very interesting to understand deeply connections of
these counterterms with the counterterms of quantum field theory for models
with continuous spin.

\subsection{Limiting Gibbs Families}

Introduce topology on the set $\mathcal{A}_{\infty }^{(0)}$ of countable
spin graphs with origin $0$, by defining the basis of open sets in it. Let $%
C_{N}$ be an arbitrary open subset of the set $\mathcal{A}_{N}^{(0)}$ of all
spin graphs with radius $N$ with respect to the origin. Then the open sets
of the basis consist of all spin graphs such that (for some $N$ and $C_{N}$)
the $N$-neighborhood of their origin belongs to $C_{N}$. As the $\sigma
-a\lg ebra$ in $\mathcal{A}_{\infty }^{(0)}$ we take the Borel $\sigma $%
-algebra, generated by these open sets.

For any spin graph $\alpha $ with root $0$ call $\gamma (\alpha ,0;N,N+d)$ $%
N $-annulus of width $d$, $\gamma (\alpha ,0;0,N)$ - $N$-internal, $\gamma
(\alpha ,0;N,\infty )$ - $N$-external parts correspondingly. Note that $N$%
-annulus and $N$-external part may be nonconnected. For given $N,d$ and some
finite graph $\gamma $ (notnecessary connected) let $\mathcal{A}%
_{N+d}^{(0)}(\gamma ,d)\subset \mathcal{A}_{N+d}^{(0)}$ be the set of all
finite connected spin graphs $\alpha $ with root $0$ and radius $N+d$ with
respect to $0$, and having $\gamma $ as the $N$-annulus of width $d$.

Let some local potential $\Phi $ of diameter $d+1$ be given.

\begin{definition}
The measure $\mu $ on $\mathcal{A}_{\infty }^{(0)}$ is called a Gibbs family
with potential $\Phi $, if for any $N,\gamma ,\Gamma $\ the conditional
distribution on the set of spin graphs $\alpha \in \mathcal{A}_{\infty
}^{(0)}$, having $N$-external part $\Gamma $ and $N$-annulus $\gamma $ of
width $d$ (this set can be naturally identified with $\mathcal{A}%
_{N+d}^{(0)}(\gamma ,d)$) coincides a.s. with the (finite) Gubbs family with
potential $\Phi $ on $\mathcal{A}_{N+d}^{(0)}(\gamma ,d)$.
\end{definition}

In particular, this conditional distribution depends only on $\gamma $, but
not on all $N$-external part $\Gamma $.

\subparagraph{Boundary conditions}

To construct pure phases, as for Gibbs fields, it is useful to have the
notion of finite Gibbs family with given boundary conditions. Boundary
conditions ($d$-boundary conditions) are given by a sequence $\nu
_{N,d}(\gamma )$ of probability measures on the set of finite (not necessary
connected) spin graphs $\gamma $, intuitively on $N$-annuli of width $d$.
Gibbs family on $\mathcal{A}_{N}^{(0)}$ with the boundary conditions $\nu
_{N+1,d}(\gamma )$ is defined as 
\begin{equation}
\mu _{N}(\alpha )=Z^{-1}(N,\nu _{N+1,d})\sum_{\xi \in \mathcal{A}%
_{N+d+1}^{(0)}(\gamma ,d;\alpha )}\int \exp (-\beta H(\xi ))d\nu
_{N+1}(\gamma ,d),\alpha \in \mathcal{A}_{N}^{(0)}
\end{equation}
\[
Z(N,\nu _{N+1,d})=\int_{\mathcal{A}_{N}^{(0)}}\left[ \sum_{\xi \in \mathcal{A%
}_{N+d+1}^{(0)}(\gamma ,d;\alpha )}\int \exp (-\beta H(\xi ))d\nu
_{N+1}(\gamma ,d)\right] d\lambda _{\mathcal{A}_{N}^{(0)}} 
\]
where $\mathcal{A}_{N+d+1}^{(0)}(\gamma ,d;\alpha )$ is the set of all spin
graphs from $\mathcal{A}_{N+d+1}^{(0)}$ with $(N+1)$-annulus $\gamma $ of
width $d$ and $N$-internal $\alpha $. Note that $\mathcal{A}%
_{N+d+1}^{(0)}(\gamma ,d;\alpha )$ is finite for any $\alpha $ and $\gamma $.

\subparagraph{Compactness}

Gibbs families on the set $\mathcal{A}_{\infty }^{(0)}$ of connected
countable spin graphs with root $0$ can be obtained as weak limits of Gibbs
families on $\mathcal{A}_{N}^{(0)}$. There can be three reasons why limiting
Gibbs states do not exist: finite Gibbs states with this potential do not
exist (see examples above), noncompactness of $S$, and also - the
distribution of finite Gibbs families can be concentrated on graphs, having
vertices with large degrees. The assumptions of the following proposition
correspond to this list, however compactness can take place under more
general conditions.

Let $\mathcal{A}_{\infty }^{(0)}(r)\subset \mathcal{A}_{\infty }^{(0)}$ be
the set of all countable spin graphs with vertices of degree not greater
than $r$.

\begin{proposition}
Let $S$ be compact. Let $\Phi $ have a finite diameter, and let $\Phi
(\gamma )=\infty $, if at least one vertex of $\gamma $ has degree $r+1$.
Let for any $N$ there exists a Gibbs family with potential $\Phi $ on $%
\mathcal{A}_{N_{{}}}^{(0)}$. Then there exists a Gibbs family with potential 
$\Phi $, with support in $\mathcal{A}_{\infty }^{(0)}(r)$.
\end{proposition}

Proof. Consider the Gibbs families $\mu _{N}$ on $\mathcal{A}_{N_{{}}}^{(0)}$%
, in fact on $\mathcal{A}_{N_{{}}}^{(0)}(r)$. Let $\tilde{\mu}_{N}$ be an
arbitrary probability measure on $\mathcal{A}_{\infty }^{(0)}(r)$\ such that
its factor measure on $\mathcal{A}_{N_{{}}}^{(0)}(r)$ coincides with $\mu
_{N}$. Then the sequence of measures $\tilde{\mu}_{N}$ on $\mathcal{A}%
_{\infty }^{(0)}(r)$ is compact, that may be proved by the standard diagonal
procedure of enumeration of all $N$-neiborhoods. It is clear that any
limiting point of the sequence $\tilde{\mu}_{N}$ is a Gibbs family with
potential $\Phi $.

\subparagraph{Gibbs families and Gibbs fields}

From the following simple result it will be clear why we use the term
''Gibbs family''. Let $\mu $ be a Gibbs family with potential $\Phi $.
Assume that the conditions of the previous proposition hold. Consider a
measurable partition of the set $\mathcal{A}_{\infty }^{(0)}$: any element $%
S_{G}$ of this partition is defined by a fixed graph $G$ (without spins) and
consists of all configurations $s_{G}$ on $G$.

\begin{proposition}
For a given graph $G$ the conditional measure on the set of configurations $%
S_{G}=\left\{ s_{G}\right\} $ is a.s. a Gibbs measure on $G$ with the same
potential $\Phi $.
\end{proposition}

Note that ''the same'' has the following obvious interpretation: if $\Lambda
_{N}$ is the $N$-neighborhood of zero in $G$ and $s_{\Lambda _{N}}$ is a
configuration on $\Lambda _{N}$, then $H(s_{\Lambda _{N}})=\sum \Phi (\gamma
)$, where the summation is over all subgraphs of the spin graph $(\Lambda
_{N},s_{\Lambda _{N}})$. Thus, any Gibbs family is a convex combination (of
a very special kind) of Gibbs measures (fields) on fixed graphs, with the
same potential. This convex combination is defined by the measure $\nu =\nu
(\mu )$ on the factor space $\mathcal{G}_{\infty }^{(0)}=\mathcal{A}_{\infty
}^{(0)}/\left\{ S_{G}\right\} $, induced by the measure $\mu $,
corresponding to the Gibbs family.

The following lemma (having will known and useful analog for Gibbs fields)
reduces the proof of the proposition to the corresponding results for finite
Gibbs families. For finite graphs this result is a direct calculation which
we omit.

\begin{lemma}
\label{lem1}Let $\mu $ be a Gibbs family on $\mathcal{A}_{\infty }^{(0)}$
with potential $\Phi $. Let $\nu _{N+1,d}(\gamma )$ be the measure on $N$%
-annuli of width $d$, induced by $\mu $. Then $\mu $ is a weak limit of the
Gibbs families on $\mathcal{A}_{N}^{(0)}$ with potential $\Phi $ and
boundary conditions $\left\{ \nu _{N+1,d}(\gamma )\right\} $.
\end{lemma}

\subsection{Empirical distributions}

One can construct probability measures on $\mathcal{A}_{\infty }^{(0)}$ in a
standard way, using Kolmogorov approach with consistent distributions on
cylindrical subsets. However this approach will not work if we would like to
construct probability measures on the set $\mathcal{A}_{\infty }$ of
equivalencs classes of countable connected spin graphs. The problem is that
there is no natural enumeration of in the countable set of vertices, and
thus it is not clear how to introduce finite-dimensional distributions. One
could say that the problem is in the absence of the coordinate system, the
reference point. That is why Kolmogorov approach cannot be directly used
here. However, some substitute of finite dimensional distributions exists.
We call this analog an empirical distribution on $\mathcal{A}_{\infty }$.

Let $S$ be finite or countable. Consider the system of numbers 
\begin{equation}
\pi =\left\{ p(\Gamma ),\Gamma \in \mathcal{A}^{(0)}\right\} ,0\leq p(\Gamma
)\leq 1,
\end{equation}
that is $\Gamma $ are finite spin graphs with root $0$. Introduce the
consistency property of these numbers for any $k=0,1,2,...$ and any fixed
spin graph $\Gamma _{k}$ of radius $k$ (with respect to $0$) 
\begin{equation}
\sum_{\Gamma _{k+1}}p(\Gamma _{k+1})=p(\Gamma _{k}),k=0,1,2,...
\end{equation}
where the sum is over all spin graphs $\Gamma _{k+1}$ of radius $k+1$ such
that $O_{k}(0)$ in $\Gamma _{k+1}$ is isomorphic to $\Gamma _{k}$. Remind
that in the latter formula it is assumed that the sum is over all
equivalence classes of spin graphs.

Assume also the normalization condition 
\begin{equation}
\sum p(\Gamma _{0})=1
\end{equation}
wher $\Gamma _{0}$ is the vertex $0$ with arbitrary spin on it.

\begin{definition}
Any such system $\pi $ is called an empirical distribution (on the set of
countable spin graphs).
\end{definition}

One can rewrite the consistency condition in terms of conditional
probabilities 
\begin{equation}
\sum_{\Gamma _{k+1}}p(\Gamma _{k+1}\mid \Gamma _{k})=1,k=0,1,...
\end{equation}
where the summation is as above and 
\begin{equation}
p(\Gamma _{k+1}\mid \Gamma _{k})=\frac{p(\Gamma _{k+1})}{p(\Gamma _{k})}
\end{equation}
Thus, we consider $\mathcal{A}^{(0)}$ as a tree, vertices of which are spin
graphs, nd there exists an edge between $\Gamma _{k+1}$ and $\Gamma _{k}$
iff $\Gamma _{k}$ is isomorphic to the $k$-neighborhood of $0$ in $\Gamma
_{k+1}$.

It is clear that the system $\pi $ defines a probability measure on $%
\mathcal{A}_{\infty }^{(0)}$. Naturalness of its interpretation as some
''measure'' on $\mathcal{A}_{\infty }$ follows from the following examples.

Examples of empirical distributions can be obtained via the following
limiting procedure, which we call an empirical limit. Let $\mu _{N}$ be a
probability measure on $\mathcal{A}_{N}$. For any $N,k$ and any spin graphs $%
\Gamma \in \mathcal{A}_{k}^{(0)},\alpha \in \mathcal{A}_{N}$ put

\begin{equation}
p^{N}(\Gamma )=\left\langle \frac{n^{N}(\alpha ,\Gamma )}{N}\right\rangle
_{\mu _{N}}
\end{equation}
wher $n^{N}(\alpha ,\Gamma )$ is the number of vertices in $\alpha $, having 
$k$-neighborhoods isomorphic to the spin graph $\Gamma $. Put $\pi
_{N}=\left\{ p^{N}(\Gamma )\right\} $. One can take measures $\mu _{N}$
equal to some Gibbs families on $\mathcal{A}_{N}$ with fixed potential. The
system $\pi _{N}$ is not an empirical distribution. However, the following
result holds.

\begin{lemma}
Let $S$ be finite and let for any $D$%
\begin{equation}
\mu _{N}(\min_{\alpha \in \mathcal{A}_{N}}D(\alpha )\leq D)\rightarrow 0
\end{equation}
for $N\rightarrow \infty $, where $D(\alpha )$ is the diameter of $\alpha $.
Then any weak point of the system $\pi _{N}$ is an empirical distribution.
\end{lemma}

Proof. Note that the probability $\mu _{N}$ that for all vertices $v$ of the
random graph one has $R_{v}(\alpha )>D$, tends to $1$ as $N\rightarrow
\infty $. that is why for any $n_{0},\delta >0$ there exists $%
N_{0}=N_{0}(n_{0},\delta )$ such that for any $N>N_{0}$ the numbers $%
p_{N}(\Gamma )$ satisfy the consistency conditions for all $\Gamma _{k}$
with $k<n_{0}$ up to $\delta $. From finiteness of $S$ the compactness
follows, that is the existence of at least one empirical distribution.

\begin{remark}
The question arises how to characterize empirical distributions which can be
obtained in this way, and in particular, what empirical distributions can be
obtained from themselves, that is the limits of numbers 
\[
p^{N}(\Gamma )=\frac{\left\langle n^{N}(\Gamma _{N},\Gamma )\right\rangle
_{\pi _{N}}}{\left\langle V(\Gamma _{N})\right\rangle _{\pi _{N}}} 
\]
where $V(\Gamma _{N})$ is the number of vertyices in $\Gamma _{N}$, $\pi
_{N} $ is the restriction of the measure $\pi $\ on $\mathcal{A}_{N}^{(0)}$.
\end{remark}

\begin{remark}
Intuitively, an empirical measure is automatically ''homogeneous'', that is
the same for any vertices. The reason is that the vertices are not
enumerated, or the graphs are not embedded to any space. Note that however,
one could easily construct ''inhomogeneous'' measures on $\mathcal{A}%
_{\infty }^{(0)}$ - it is sufficient to take $\Phi $ depending on the
distance from $0$.
\end{remark}

\begin{remark}
Another approach to empirical distributions is possible, based not on
superlocality but on locality, when we ''do not know'' all neighborhood of a
vertex. Let, for example, two graphs $\gamma \subset \Gamma $ be given such
that in $\Gamma $ there is no other subgraph isomorphic to $\gamma $. Then
one can look for the limits of the following fractions (conditional
empirical distributions) 
\[
\left\langle \frac{n^{N}(\alpha ,\Gamma )}{n^{N}(\alpha ,\gamma )}%
\right\rangle _{\mu _{N}} 
\]
\end{remark}

\subsubsection{Empirical Gibbs families}

Let us show that sometimes limiting Gibbs families on $\mathcal{A}_{\infty
}^{(0)}$\ can be obtained via the limit not in the radius tending to
infinity but in the number of vertices. Let $\mathcal{B}_{N,p}^{(0)}$ be the
set of spin graphs with $N$ vertices, root $0$, and such that each vertex
has the degree not greater than $p$. Consider the finite Gibbs family $\mu
_{N}$ on $\mathcal{B}_{N,p}^{(0)}$ with the local potential $\Phi $.

\begin{lemma}
Any weak limiting point of the sequence $\mu _{N}$ is a limiting Gibbs
family on $\mathcal{A}_{\infty }^{(0)}$ with potential $\Phi $.
\end{lemma}

Proof. Let us fix $n,d$ and some $\alpha \in \mathcal{A}_{n+d+1,p}^{(0)}$
with $n$-annulus $\gamma $ of width $d+1$. Then for any sufficiently large $%
N $ there exists $\beta \in \mathcal{B}_{N,p}^{(0)}$ such that $%
O_{n+d+1}(\beta ,0)$ is isomorphic to $\alpha $. For any such $\beta $ the
conditional distribution $O_{n}(\alpha ,0)$ under given $\gamma $ is Gibbs
with potential $\Phi $.

An empirical distribution $\pi $ on $\mathcal{A}_{\infty }$ is called Gibbs
if ois obtained as the empirical limit of finite Gibbs families $\pi _{N}$
on $\mathcal{A}_{N}$ \ with potential $\Phi $.

To justify this definition, let us show that under some conditions an
empirical distribution $\pi $ on $\mathcal{A}_{\infty }$ is a Gibbs family
as a measure $\mathcal{A}_{\infty }^{(0)}$. Assume that the empirical
distribution $\pi $ on $\mathcal{A}_{\infty }$ is Gibbs with potential $\Phi 
$, equal to $\infty $, if the degree of at least one vertex is greater than $%
p$, and that the probability $\pi _{N}$ of the event that the automorphism
group of the graph $\alpha \in \mathcal{A}_{N}$ is trivial, tends to $1$ as $%
N\rightarrow \infty $. Then $\pi $ is a Gibbs family as the measure on $%
\mathcal{A}_{\infty }^{(0)}$. In fact, is a graph has trivial automorphism
group the, one can fix root at any vertex, what gives different rooted
graphs. Then we have Gibbs families on $\mathcal{B}_{N,p}^{(0)}$, and the
previous lemma we get a limiting Gibbs family on $\mathcal{A}_{\infty
}^{(0)} $.

\subsubsection{Empirical distributions in planar quantum gravity}

One should note that problem of existence of limiting correlation functions,
normally ignored in physics, is far from being trivial. Besides its
mathemtical interest, there another its aspect, which could have some
interest for physics as well. In physical discrete quantum gravity, where
normally grand canonical ensemble is used, the thermodynamic limit has
physical sense only in some critical point, which defines the boundary of
the region where the series defining the grand partition function,
converges. This critical point moreover, strongly depends on geometric
details even for the fixed topology. When the topology is not fixed, the
corresponding series diverges for all parameters, see below, even in
dimension 2, because of the factorial asymptotics of entropy for canonical
partition functions. Canonical ensemble have no such difficulties but the
question arises of how limiting correlations depend on the choice of the
canonical ensemble.

Here we give popular examples of empirical Gibbs families. In physics the
pure (without matter fields) planar quantum gravity (or string in dimension $%
0$) is defined by the grand partition function 
\[
P(T)=Z^{-1}\exp (-\mu F(T)),Z=\sum_{T\in \mathcal{A}}\exp (-\mu
F(T))=\sum_{N}C(N)\exp (-\mu N) 
\]
where the first sum is over all triangulations $T$ of the sphere with $N$\
triangles, belonging to some class $\mathcal{A}$ \ of triangulations. The
exact definition of the class $\mathcal{A}$\ is not essnetial, see\ \cite{m3}%
, but leads however to different empirical distributions for different $%
\mathcal{A}$. $C(N)=C_{\mathcal{A}}(N)$ is called the entropy and equals the
number of such triangulations. It is known that 
\[
C(N)\sim aN^{-\frac{7}{2}}c^{N} 
\]

One could also consider $N$ to be the number of vertices of the dual graph,
which is a $3$-regular graph, imbedded in the two-dimensional sphere. Grand
canonical ensemble is normally considered for $\mu \rightarrow \mu
_{cr}+0=-\log c$. In the canonical ensemble one does not need the parameter $%
\mu $.

Let us consider a triangulation with the root vertex $0$ and fixed
neighborhood $O_{d}(0)=\Gamma _{d}$ of radius $d$. Let $p^{N}(\Gamma _{d})$
be its probability in our canonical ensemble.

\begin{theorem}
There exists the limit $\lim p^{N}(\Gamma _{d})=\pi (\Gamma _{d})$, which
defines an empirical distribution.
\end{theorem}

We give the proof which uses explicit formulae.. Let $\Gamma _{d}$ have $u$
triangles and $k$ edges on the boundary of the neighborhood that is (that is
edges with both vertices on the $d$-annulus). Let $r(d,k,u)$ be the number
of such neighborhoods. The external part is a triangulation of a disk with $%
N-u$ triangles and $k$ boundary edges. The number $D(N-u,k)$ of such
triangulations (this follows from results by Tutte, see \cite{m3}) has the
following asymptotics as $N\rightarrow \infty $ 
\begin{equation}
D(N-u,k)\sim \phi (k)N^{-\frac{5}{2}}c^{N-u}
\end{equation}
That is why the probability of $\Gamma _{d}$ (at vertex $0$) equals 
\begin{equation}
P^{N}(\Gamma _{d})\sim \frac{\phi (k)N^{-\frac{5}{2}}c^{N-u}}{%
\sum_{u,k}r(d,k,u)\phi (k)N^{-\frac{5}{2}}c^{N-u}}=\frac{\phi (k)c^{-u}}{%
\sum_{u,k}r(d,k,u)\phi (k)c^{-u}}
\end{equation}
The functions $\phi ,r$ and the constant $c$ are known explicitely, but it
is not important for us. As almost all graphs have trivial automorphism
group, it follows that the empirical limit is the same.

Note that if instead of the sphere we would consider a surface of genus $%
\rho $, we would get another empirical distribution, depending on $\rho $
and $\mathcal{A}$.

Is this empirical distribution (planar) a Gibbs empirical distribution?\ Two
dimensional triangulations can be selected with the hard core potential $%
\Phi $ such that $\Phi (O_{1}(v))=\infty $, if $O_{1}(v)\neq g_{k,k},k\geq 2$%
. However this does not distinguish sphere and other two dimensional
surfaces. That is why we have two possibilities. The firs one is to get
triangulations of the sphere as a pure phase of some Gibbs family. We shall
construct such an example later. Second one is to consider nonlocal
potentials (of infinite radius). The last characterization is well-known in
graph theory. Consider the potential $\Phi (K_{5})=\Phi (K_{3,3})=\infty $,
if $K_{5}$ is a (not regular) subgraph, homeomorphic to the complete graph
with 5 vertices, and $K_{3,3}$ is a (not regular) subgraph, homeomorphic to
the graph with 6 vertices $1,2,3,4,5,6$ and all edges $(i,j),i=1,2,3,j=4,5,6$
(by Pontriagin-Kuratowski theorem this selects planar graphs). Put $\Phi =0$
otherwise. Thus the planarity one should consider as some apriori condition
of global character. Note however, that without topological restrictions we
have an empirical Gibbs distribution.

\section{Entropy}

Famous van Hove theorem in statistical physics says that the partition
function grows exponentially with the volume. In quantum gravity this result
concerning the number of triangulation (we call it here entropy, not its
logarithm) was a subject of many papers, see \cite{amcama}. Only the case
when the topology is fixed or subject to strong restrictions, was considered
(see \cite{boulatov}). However, it is easy to see that the partition
function grows faster than exponentially even in the two dimensional case if
the genus is not fixed. That is why the grand canonical ensemble, as it is
presented in physical papers, does not exist if canonical partition
functions have superexponential growth. Double scaling limit is an
(nonrigorous) attempt to overcome this difficulty.

Thus, the question about the entropy growth in the general is obvious
interest. Below we shall give examples of subexponential (power) growth,
where however there will be a phase transition. At the same time the number
of \ $p$-regular graphs with $N$ vertices is of order $N^{\frac{pN}{2}}$.
Note also that appending of the spin to a Gibbs graph could influence only
on the toipologies which are typical (that is add to maximal entropy only
exponential factor). That is it is useful to know rough (logarithmic)
asymptotics of the entropy as well.

\subsection{Arbitrary genus in two dimensions}

The asymptotics of the number of triangulations of a surface of fixed genus $%
g$ is 
\begin{equation}
C(N,g)\sim f(g)N^{ag+b}c^{N},a=\frac{5}{2},b=-\frac{5}{2}-1  \label{rod}
\end{equation}
where $c$ does not depend on $g$. I am not aware of the formally written
proof of this result, but it seems reasonable as it is confirmed by the
results of \cite{bender} for other classes of maps, obtained by a direct by
laborous development of Tutte's methods. Note that another method of
reducing the problem to a matrix model, though gives the same result but
cannot be considered as a rigorous proof. Much more difficult is the case
when the genus is not fixed. We give some first results in this direction,
see also \cite{KriMal} where stronger results are obtained.

We give necessary definiyions from \cite{wale}. A map is a triple $(S,G,\phi
)$, where $S$ is a closed oriented connected compact two dimensional
manifold, $G$ is a connected graph, $\phi $ is a homeomorphic embedding of $%
G $ in $S$, such that the complement to $\phi G$ is the union of connected
open sets, each homeomorphic to an open disk. A rooted map ia a map with one
specified edge-end (that is an edge with one of its vertices) of $G$. Two
rooted maps are (combinatorially) equiavalent if there exists an orientation
respecting homeomorphism of $S$, which transforms the vertices to the
vertices, the edges to the edges, the root to the root. A combinatotrial map
(further on simply a map) is an equivalence class of maps. There exists an
equivalent combinatorial definition of maps as a cyclic (ribbon, fat) graph.
Cyclic graph is a graph with given cyclic order of edge-ends in each vertex.
It is based on the following theorem, coming back to Edmonds.

\begin{theorem}
For any connected cyclic graph there is a topologically unique imbedding in
a surface so that the order of edge-ends becomes the clockwise ordering and
the complement of the graph on the surface is a set of disks.
\end{theorem}

This theorem shows why the problem becomes solvable: it is reduced to
simpler graph counting problem. It is interesting to note that for 3-dim
case the counting of complexes can be reduced to graphs only for
pseudomanifiolds.

Let $E$ be the set of edges, then $2E=E\times \left\{ -1,1\right\} $ - the
set of edge-ends of the graph, vertices are defined by a partition of $2E$.
Each block of the partition has a cyclic order of its elements, thus
defining a permution $P$ on $2E$. Another permutation, minus $(-)$, changes
edge-ends of the same edge, i. e. $1$ and $-1$. A face is defined by a
sequence of edge-ends $e,(-)e,P(-)e,(-)P(-)e,...$. Thus, cycles of $P$ are
vertices, cycles of $(-)$ are edges and faces correspond to cycles of the
permutation $P(-)$. This allows to give the following definition.

\begin{definition}
A combinatorial map is defined by a triple $(2E,P,(-))$, consisting of the
set $2E$ with even number of elements and two its permutations such that the
group, generated by them is transitive on this set (this corresponds to
connectedness of $G$). It is assumed also, that the permutation $(-)$
consists of nonintersecting cycles of order $2$. The root is a specified
element of $2E$.
\end{definition}

Let $F_{b,p}(g)$ be the number of rooted maps with $p+1$ vertices and $b+p$
edges on the surface of genus $g$. Put $F_{b,p}=\sum_{g}F_{b,p}(g)$.

\begin{theorem}
Let $p(b)$ be a nondecreasing sequence of integers and such that $\frac{p(b)%
}{b}\rightarrow \alpha $ for some constant $0<\alpha <\infty $. Then 
\[
b^{-1}\ln \frac{F_{b,p(b)}}{b!}\rightarrow c=c(\alpha ),0<c(\alpha )<\infty 
\]
\end{theorem}

In\ \cite{wale} the following recurrent equations for $F_{b,p}$ are obtained 
\begin{equation}
F_{b,p}=\sum_{j=0}^{p-1}%
\sum_{k=0}^{b}F_{k,j}F_{b-k,p-j-1}+(2(b+p)-1)F_{b-1,p},b,p\geq 0
\label{rec1}
\end{equation}
except the case $b=p=0$. Moreover 
\[
F_{0,0}=1,F_{-1,p}=F_{b,-1}=0 
\]
One should separately consider the cases $b=0$ or $p=0$. The case $b=p=0$
corresponds to the imbedding of one vertex to the sphere. The cases $p\geq
1,b=0$ correspond to a tree, imbedded also to a sphere, $F_{0,p}=\frac{(2p)!%
}{p!(p+1)!}$. $F_{b,0},b\geq 1,$ are equal to $(2b-1)(2b-3)...=\frac{(2b)!}{%
b!2^{b}}$, the number of partitions of the set $\left\{ 1,2,...,2b\right\} $
in pairs.

\begin{lemma}
The following estimates hold 
\[
c_{1}^{b+p}<G_{b,p}=\frac{F_{b,p}}{b!}<c_{2}^{b+p},0<c_{1}<c_{2}<\infty 
\]
\end{lemma}

Proof. We have from (\ref{rec1}) 
\begin{equation}
G_{b,p}=\sum_{j=0}^{p-1}\sum_{k=0}^{b}\frac{k!(b-k)!}{b!}%
G_{k,j}G_{b-k,p-j-1}+b^{-1}(2(b+p)-1)G_{b-1,p},b,p\geq 1  \label{rec2}
\end{equation}
One can get a lower bound from the following minorizing recurrent relations 
\[
E_{b,p}>2E_{b-1,p},b\geq 1,p\geq 1,E_{0,p}=G_{0,p}=\frac{(2p)!}{p!(p+1)!} 
\]
To get the upper bound we need some techniques, further used in \cite{KriMal}%
, of trees expansion, which correponds to the complete iteration of the
following majorizing recurrence 
\begin{equation}
G_{b,p}=\sum_{j=0}^{p-1}\sum_{k=0}^{b}\frac{k!(b-k)!}{b!}%
G_{k,j}G_{b-k,p-j-1}+(2+\frac{2p}{b})G_{b-1,p}  \label{rec3}
\end{equation}
with the same boundary conditions $G_{0,0}=1,G_{b,p}=0$, if either $p=-1$ or 
$b=-1$. We shall consider pairs $(T,\phi )$, where $T$ \ is a finite rooted
tree, $\phi $ is a mapping of the set vertices $V(T)$ of the tree to the
positive quadrant of the plane, denote $\phi (v)=(b=b(v),p=p(v))$ for $v\in
V(T)$. Let $\Phi (b,p)$ be the set of all pairs $(T,\phi )$, where the root
corresponds to the point $(b,p)$. Any such tree corresponds to some steps of
iteration of the recurrence (\ref{rec3}). New vertex of the tree (we say
that the new vertex is under some old vertex) corresponds to some iteration
step: either we choose one of the terms in the right hand side of (\ref{rec3}%
) or we stop for a given vertex. Any vertex $v=(b,p)$ of the tree (except
the root) has one ingoing edge and $0,1$ or $2$ outgoing edges, in the
latter case the vertex is called binary, then two other vertices of these
two edges are in the points $(k,j)$ and $(b-k,p-j-1)$. Then we prescribe to
this vertex the factor $f(v)=\frac{k!(b-k)!}{b!}$. If there is only one
outgoing edge then the vertex is called unary, in this case other vertex is
in the point $(b-1,p)$. Then the vertex $v=(b,p)$ has a factor $f(v)=2+2%
\frac{p}{b}$. The vertex $v=(b,p)$, from which there are no outgoing edges
is called, is called final and has a factor $f(v)=G_{b,p}$. Contribution of
the tree $T$ we call the following product 
\[
f(T)=\prod_{v\in V(T)}f(v) 
\]
The tree is called final if to each its final vertex corresponds the point $%
(0,0)$.

\begin{lemma}
The following expansion holds 
\[
G_{b,p}=\sum_{T}f(T) 
\]
where the summation is over all trees.
\end{lemma}

The proof is obtained by complete iteration of the recurrent equation (\ref
{rec3}), where we stop only when there will not be any final vertices in the
points different from $(0,0)$.

Note that the number of binary vertices in any tree with the root in $(b,p)$
is not greater than $p$. In fact, put $p(T)=\sum p(v)$, where the sum is
over all final vertices $v$ of the tree $T$. \ Let the tree $T^{\prime }$ be
obtained from $T$ by appending two edges to some of some final vertex of the
tree $T$. Then $p(T^{\prime })=p(T)-1$. If $T^{\prime }$ is obtained from $T$
by appending one edge to some of some final vertex of the tree $T$, then $%
p(T^{\prime })=p(T)$. The result follows.

\begin{lemma}
The numeber $K_{b,p}$ of pairs $(T,\phi )$, where $T$ are complete trees
with the root in $(b,p)$, does not exceed $C^{b+p}$ for some constant $C$.
\end{lemma}

Proof. We have the following recurrent equation 
\[
K_{b,p}=\sum_{j=0}^{p-1}\sum_{k=0}^{b}K_{k,j}K_{b-k,p-j-1}+K_{b-1,p} 
\]
The equation for the generating function 
\[
K=K(x,y)=\sum K_{b,p}x^{b}y^{p} 
\]
is 
\[
K=xK^{2}+yK+1 
\]
It follows that 
\[
K=-\frac{y-1}{2x}+\frac{y-1}{2x}\sqrt{1-\frac{4x}{(y-1)^{2}}} 
\]
From analyticity of $K$ in the point $x=y=0$ we get the result.

By the previous lemma it is sufficient to prove that for any pair $(T,\phi )$
the contribution does not exceed $C^{b+p}$ for some constant $C$ not
depending on the pair. Some difficulty provides the estimate of the product
of $\frac{2p}{b}$ in all unary pairs. Let us prove by induction that the
number of unary vertices is not greater than $b$. Start with the root vertex 
$(b,p)$. Let between the root and the next binary vertex there is exactly $%
b_{1}$ unary vertices. The binary vertex $(b-b_{1},p)$ creates two other
vertices $(b^{\prime },p^{\prime })$ and $(b^{\prime \prime },p^{\prime
\prime })$ with $b^{\prime }+b^{\prime \prime }=b-b_{1},p^{\prime
}+p^{\prime \prime }=p-1$. By induction under these two vertices there is
not more than $b^{\prime }+b^{\prime \prime }$ unary vertices. That is why
the general number of binary vertices does not exceed $b^{\prime }+b^{\prime
\prime }+b_{1}=b$.

There is a set $V^{\prime }$ of vertices with the following properties: 1)
no vertex of $V^{\prime }$ is under some other vertex of $V^{\prime }$, 2)
any vertex $(b^{\prime \prime },p^{\prime \prime })$, which is not under a
vertex of $V^{\prime }$, has the property that $b^{\prime \prime }\geq
2p^{\prime \prime }$, 3) for any vertex $(b^{\prime },p^{\prime })$ from $%
V^{\prime }$ we have $b^{\prime }<2p^{\prime }$.

The contribution of unary vertices of type 2) we shall estimate as $1$. Take
now some vertex of $V^{\prime }$ with $b^{\prime }<\frac{p^{\prime }}{\ln
p^{\prime }}$. General number of unary vertices under it does not exceed $%
b^{\prime }$ and thus their contribution does not exceed $(2p^{\prime })^{%
\frac{p^{\prime }}{\ln p^{\prime }}}<C^{p^{\prime }}$.

Let now $\frac{p}{\ln p}\leq b<2p$ for some vertex $(b,p)$. Denote $%
g(T)=b!f(T)$, if $(b,p)$ is the root of $T$. If the root is a unary vertex
wuth contribution $\frac{2p}{b}$, then $g(T)<2pg(T^{\prime })$, as the
vertex under it is $(b-1,p)$. If the root is a binary vertex with
contribution $\frac{k!(b-k)!}{b!}$, then $g(T)=g(T_{1})g(T_{2})$, where $%
T_{i}$ are trees with roots $(k,j)$ and $(b-k,p-j-1)$ correspondingly, for
some $j$. It follows that $g(T)<(2p)^{b}$ and thus 
\[
f(T)<\frac{(2p)^{b}}{b!}<e^{2p} 
\]
Note that $\sum_{v^{\prime }\in V^{\prime }}p(v^{\prime })\leq p$, the reult
follows.

Let us finish the proof of the theorem. Consider the function of one complex
variable 
\[
f(z)=\sum_{b=0}^{\infty }z^{b+p(b)}G_{b,p(b)} 
\]
By the previous lemma this series has finite radius of convergence, say $R$.
Then for any $\varepsilon >0$ and $b=b(\varepsilon )$ sufficiently large, we
have 
\[
(R^{-1}-\varepsilon )^{b}<G_{b,p(b)}<(R^{-1}+\varepsilon )^{b} 
\]
The result follows.

It is useful to present the result of the theorem in a different way. Let $%
\rho (M)$ be the minimal genus where $M$ can be inmedded and let $\frak{M}%
(b,p)$ be the set of maps with parameters $b,p$. Put $\rho (b,p)=\max_{M\in 
\frak{M}(b,p)}\rho (M)$. Then the theorem says that $\frac{\left| \frak{M}%
(b,p)\right| }{(2\rho (b,p))!}$ has an exponential growth.

Otherwise speaking, the factorial factor has the order $(-\chi _{\min })!$,
where $\chi _{\min }$ is the minimum of Euler characteristics of possible
manifolds.

This is seen from the formula 
\[
-\chi =2\rho -2=-V+L-F=b-1-F=-z(P,-)+b-1 
\]
where $z(P,-)$ is the number of orbits of the group generated by both
permutations. It is not difficult to see that for given orbits of $P$ one
can always choose $-$ so that the orbits of $(P,-)$ were as long as
possible, that is $z(P,-)$\ will give the contribution much less than $b$.

This result allows to suggest the following reasonable canonical
distribution on the set $\mathcal{A}(b)$ of all maps $M$ with fixed $b=-\chi
_{\min }$ 
\[
P_{b}(M)=Z_{b}^{-1}\exp (-\mu p(M)-\lambda \rho (M))=Z_{b}^{-1}\exp (-\mu
p(M)+\frac{\lambda }{2}F(M)) 
\]
Unfortunately, there are no results concerning asymptotics of this partition
function, but see \cite{KriMal}.

\subsection{Dimension 3, 4}

In quantuml gravity for a fixed triangulated $d$-dimensional manifold \ $M$
(simplicial complex) the potential is linear in the numbers $N_{i}(T)$ of $i$%
-dimensional simplices of the triangulation $T,i=0,...,d,$ that is the
pratition functions is formally 
\[
Z=\sum_{T}\exp (-\sum_{i}k_{i}N_{i}(T))=\sum_{\left\{ N_{i}\right\} }\exp
(-\sum_{i}k_{i}N_{i})C(\left\{ N_{i}\right\} ) 
\]
where $C(\left\{ N_{i}\right\} )$ is the number of triangulation with given
numbers $C(\left\{ N_{i}\right\} )$. From Dehn-Sommerfeld relations (see 
\cite{amcama}) it follows that only two of $N_{i}(T)$\ are linearly
independent. That is why the distribution can be written as follows 
\[
\sum_{T}\exp (-\lambda _{d}N_{d}(T)+\lambda _{d-2}N_{d-2}(T)) 
\]
where the sum is over all triangulations $T$ of the given manifold. Moreover
the latter formula corresponds to the Hilbert-Eisnstein action 
\[
S=\frac{1}{16\pi G}\int dx\sqrt{g}(2\Lambda -R) 
\]
where $R$\ is the scalar curvature, $\Lambda $\ is the cosmological
constant, $G$\ is the Newton constant. For the convergence of the latter
series at least for some parameters one needs exponential estimates on $%
C(\left\{ N_{i}\right\} )$. Discussion of necessary exponential estimates
see in \cite{amcama}. D. Boulatov \cite{boulatov} gives the following
result. Let $C(N,\mathbf{M})$ be the number of three dimensional
triangulations with $N$ simplices of the three dimensional manifold $\mathbf{%
M}$. Then 
\[
\sum C(N,\mathbf{M})<C\exp (\gamma N) 
\]
where the sum is over all $\mathbf{M}$ with fixed homology group $H_{1}(%
\mathbf{M},Z),\gamma =\gamma (H_{1}(\mathbf{M},Z))$.

In the general case (arbitrary topology) the growth is not exponential, and
Gibbs families provide the unique possibility to introduce a probability
distribution.

To each $d$-dimansional simplicial complex there corresponds a dual graph,
vertices of which are in one-to-one correspondence with $d$-simplices. Each
vertex of this graph has $d$ edge-ends (legs), which correpond to to $d$ of
its $(d-1)$-dimensional faces. Coupling of edge-ends corresponds to glueing
of the correponding faces with linear mappings. In dimansion 2 \ one gets a
manifold for any way of coupling. In larger dimensions we get
pseudomanifolds with probability tending to $1$. Moreover, in the
thermodynamic limit this pseudomanifold does not have any vertex with a
neighborhood isomorphic to a ball. This follows from results of the section
4.4

The condition selecting manifolds among the pseudomanifolds is local and can
be given by a superlocal potential (with hard core). However restrictions on
topology are not local in general. However, below we consider examples where
restrictions on topology will be a corollary of the Gibbs property.

\section{Phase Transitions}

Gibbs family (on $\mathcal{A}_{\infty }^{(0)}$\ or on $\mathcal{A}_{\infty }$%
) with potential $\Phi $ is called a pure Gibbs family if it is not a convex
combination of other Gibbas families with the same potential. Thus,
nonuniqueness of for Gibbs families means that there at least two pure Gibbs
families with the same potential $\Phi $. Pure Gibbs state can be can be of
two kinds: either it has the support on the set configurations on one fixed
graph or it has the support on several or even continuum different graphs.
In the fisrt case it reduces of course to a Gibbs field.

Accordingly to this nonuniqueness can be of quite different nature. Firstly,
on each graph of the support of the Gibbs family the pure phases can differ
with the structure of typical configurations (this is a phase transition
inherited from Gibbs fields a fixed graph). Secondly, pure phases can differ
with the typical structure of the graph itself (metrical or topological
phase transition). Heuristically speaking, topological phase transition
takes place when there are at least two Gibbs families with non-intersecting
supports, and these supports have typical graphs with quite different
topological properties (or quite different local structure, defining the
topology).

Description of all phases is sufficiently complicated even for simple
examples. It is simpler to select some phases, we do this below for some
examples. The following remark will be useful: if a pure Gibbs field is a
Gibbs family then it is a pure Gibbs family.

\subsection{Mean-field model}

In discrete quantum gravity there is a simple model \cite{bibujo}, which is
explicitely (and rigorously) solvable, where different phases roughly
correspond to numerical modelling of more complicated models with spin. This
is an urn model with $M=\lambda N$ balls in $N$ urns. It can be interpreted
as a random graph, if one calls urns vertices and balls edge-ends. After
this edge-ends are coupled in pairs. Thus, there are $N$ labelled
(enumerated) vertices $i=1,...,N$, with $q_{i}\geq 1$ edge-ends of a vertex $%
i$. Canonical ensemble is given by the partition function 
\[
Z_{N}=\sum_{\left\{ q_{i}\right\} :\sum q_{i}=M}\exp (-\sum_{i=1}^{N}\beta
\Phi (q_{i})) 
\]
Note that in our terminology this is a model with a local potential (not
stable, as it was shown above) and nonlocal restriction $\sum
q_{i}=M=\lambda N$ of the mean field type. This restriction restores
stability for such potentials. Note that the factorial factor $A(M)$ in the
asymptotics of the partition function, which is responsible for random
coupling of edge-ends can be factorized, after what the asymptotics looks as
follows 
\[
Z_{N}=cN^{\alpha }\exp (\gamma N) 
\]

For sufficiently wide class of potentials, including $\Phi (q)=\log q$,
there are two phases with a phase transition in $\lambda $, for fixed $\beta 
$. One phase (elongated, fluid) for small densities, where the balls are
uniformly distributed in the urns, the secodn one (crumpled, Bose-Einstein)
phase, where in one only vertex the number of balls is about $M$ balls. With
some modification of the model one can get a mixed phase (condensed,
crinkled), where there is one vertex with $cM,c<1,$ balls, the other urns
are uniformly filled up.

Numerical modelling, see \cite{thorl}, shows the appearance of similar
phases in two-dimensional gravity with spins. That is why various
explanations and analogs are of interest. As we already said, in fact this
is not more than a smoothed transition between stability and instability of
the potential, the smoothing being realised by a mean field type restriction.

In statistical physics \ the analog is the Bose-Einstein condensation. In
informatics the choice\ $\Phi =cq$ gives the stationary distribution in
Jackson networks in the mean field model, see \cite{maya}, where quite
similar phase transition were observed. All classical theory of random
graphs deals in fact with the chemical potential with the same mean field
type restriction, see \cite{bol}.

\subsection{Phase transition with boundaries}

We consider triangulations $T$ of a two-dimensional sphere with $k$ holes,
called quasitriangulations in \cite{goujac} and triagulation in \cite{m3}.
These maps consist of vertices, edges and faces-triangles (that is having
three smooth sides), but are not assumed to be an abstract simplicial
complex. For example, several edges can connect two vertices. Let $V(T)$ ($%
L(T)$) be the set of vertices (edges) in $T$, $L(T)$ includes the boundary
edges.

Let $T(\rho ,N,k;m_{1},...,m_{k})$ be the set of all such (combinatorial)
triangulations with $N$ triangles and $m_{i}\geq 2$ edges on the boundary of
the hole $i=1,...,k$. Assume that $m_{i}\geq 2$. Let $C(\rho
,N;m_{1},...,m_{k})$ be the number of such triangulations. For fixed $\rho
,k $ and given $N$ define the distribution $\mu _{\rho ,N,k}$ (canonical
ensemble) on the set 
\[
T(\rho ,N,k)=\cup _{m_{1},...,m_{k}}T(\rho ,N,k;m_{1},...,m_{k}) 
\]
of triangulations. One can introduce $T(\rho ,N,k)$\ in four equivalent
(easy to verify) ways:

\begin{itemize}
\item  interaction is proportional to the general number of edges, that is 
\[
\mu _{\rho ,N,k}(T)=Z_{N}^{-1}\exp (-\mu _{1}L(T)) 
\]
where $L(T)$ is the number of edges of $T$, including all boundary edges;

\item  interaction is proportional to the sum of of degrees of vertices 
\[
\exp (-\mu _{1}L(T))=\exp (-\frac{\mu _{1}}{2}\sum_{v\in V(T)}\deg v) 
\]
Thus 
\[
Z_{N}=Z_{N}(\rho ,k)=\sum_{(m_{1},...,m_{k})}\sum_{T\in T(\rho
,N;m_{1},...,m_{k})}\exp (-\frac{\mu _{1}}{2}\sum_{v\in V(T)}\deg v) 
\]
Using a duscrete analog of the Gauss-Bonnet theorem, one can show that this
is a discrete analog of Einstein-Hilbert action. Note that this is a
particular case (with parameters $t_{q}=t$) of the interaction , considered
in \cite{kastwy} 
\[
\prod_{q>2}t_{q}^{n(q,T)} 
\]
where $n(q,T)$ is the number of vertices of degree $q$.

\item  interaction is proportional to the number of boundary edges.\ For any 
$\rho ,k$\ define probability distributions on the set of vectors $%
(m_{1},...,m_{k})$ 
\[
P_{N,\rho ,k}(m_{1},...,m_{k})=\sum_{T\in T(\rho ,N;m_{1},...,m_{k})}\mu
_{\rho ,N,k}(T)= 
\]
\[
=\Theta _{N}^{-1}(\rho ,k)\exp (-\mu _{1}\frac{\sum_{i=1}^{k}m_{i}}{2}%
)C(\rho ,N;m_{1},...,m_{k}) 
\]
We used the formula 
\[
L(T)=\frac{3N-\sum m_{i}}{2}+\sum m_{i}=\frac{3N}{2}+\frac{\sum m_{i}}{2} 
\]
Then 
\[
\Theta _{N}(\rho ,k)=\sum_{m_{1},...,m_{k}=2}^{\infty }\exp (-\mu _{1}\frac{%
\sum m_{i}}{2})C(\rho ,N;m_{1},...,m_{k}) 
\]

\item  introduce fictitious spins $\sigma _{v}$ in the vertices $v$ of the
triangulation, taking values in any compact set, and assume that the
potential of the Gibbs family is 
\[
f(\sigma _{v},\sigma _{v^{\prime }})\equiv 1 
\]
for any two neighboring vertices $v,v^{\prime }$.
\end{itemize}

Consider first the case $k=1,\rho =0$. Moreover assume that on the boundary
one vertex and its incident edge are specified, defining thus the origin and
the orientation. Denote the class of such triangulations $T_{0}(N,m)$, and
let $C_{0}(N,m)$ be their number. Then the distribution is 
\[
P_{0,N}(T)=Z_{0,N}^{-1}\exp (-\frac{\mu _{1}}{2}\sum_{v\in V(T)}\deg v) 
\]
We will observe phase transitions with respect to parameter $\mu _{1}$.
Denote $\mu _{1,cr}=\log 12$. Let $0<\beta _{0}=\beta _{0}(\mu _{1})<1$ be
such that 
\[
\frac{(1+\frac{4\beta _{0}}{3(1-\beta _{0})})}{(1+\frac{2\beta _{0}}{1-\beta
_{0}})^{2}}\exp (-\mu _{1}+\log 12)=1 
\]

\begin{theorem}
The free energy $\lim_{N}\frac{1}{N}\log Z_{0,N}=F$ is equal to $-\frac{3}{2}%
\mu _{1}+c,c=3\sqrt{\frac{3}{2}}$, if $\mu _{1}>\mu _{1,cr}$ and to 
\[
-\frac{3}{2}\mu _{1}+c+\beta _{0}(-\mu _{1}+\log 12)+\int_{0}^{\beta
_{0}}\log (\frac{(1+\frac{4\beta }{3(1-\beta )})}{(1+\frac{2\beta }{1-\beta }%
)^{2}}d\beta 
\]
\ if $\mu _{1}<\mu _{1,cr}$.
\end{theorem}

Note that if $\mu _{1}\rightarrow \mu _{1,cr}$ then $\beta _{0}\rightarrow 0$%
.

Let $m(N)$ be the random length of the boundary when $N$ is fixed. Its
probability can be written as, using $\left| L(T)\right| =\frac{3N}{2}+\frac{%
m}{2}$,

\[
P_{0,N}(m(N)=m)=\Theta _{0,N}^{-1}\exp (-\mu _{1}\frac{m}{2}%
)C_{0}(N,m),\Theta _{0,N}=\sum_{m}\exp (-\mu _{1}\frac{m}{2})C_{0}(N,m) 
\]

\begin{theorem}
There are 3 phases, where the distribution of $m(N)$ has quite different
asymptotical behaviour:

\begin{itemize}
\item  Subcritical region, that is $12\exp (-\mu _{1})<1$: $m(N)=O(1)$, more
exactly the distribution of $m(N)$ has a limit $\lim_{N}P_{N}(m(N)=m)=p_{m}$
for fixed $m$ as $N\rightarrow \infty $. Thus the hole becomes neglectable
with respect to $N$.

\item  Supercritical region (elongated phase), that is $12\exp (-\mu _{1})>1$%
. Here the boundary length is of order $O(N)$. More exactly there exists $%
\varepsilon >0$ such that $\lim P_{0,N}(\frac{m_{N}}{N}>\varepsilon )=1$.

\item  In the critical point, that is when $12\exp (-\mu _{1})=1$, the
boundary length is of order $\sqrt{N}$. The exact statement is that the
distribution of $\frac{m_{N}}{\sqrt{N}}$ converges in probability.
\end{itemize}
\end{theorem}

Proof. We use the following formula from \cite{goujac}, where $N=m+2j$, 
\[
C_{0}(N,m)=\frac{2^{j+2}(2m+3j-1)!(2m-3)!}{(j+1)!(2m+2j)!((m-2)!)^{2}} 
\]
By direct calculation we get, taking into account that $N\rightarrow
N,m\rightarrow m+2$ corresponds to $j\rightarrow j-1,m\rightarrow m+2$ 
\[
\frac{P_{0,N}(m+2)}{P_{0,N}(m)}=f(N,m)=\frac{C_{0}(N,m+2)\exp (-\frac{\mu
_{1}}{2}(m+2))}{C_{0}(N,m)\exp (-\frac{\mu _{1}}{2}m)}= 
\]
\begin{equation}
=\exp (-\mu _{1}+\log 12)\frac{(1+\frac{2}{N-m})(1+\frac{4m}{3(N-m)})}{(1+%
\frac{2m}{N-m}+\frac{2}{N-m})(1+\frac{2m}{N-m}+\frac{1}{N-m})}\frac{(1-\frac{%
1}{4m^{2}})}{(1-\frac{1}{m})}  \label{main}
\end{equation}
In the subcritical case for fixed $m$ and $N\rightarrow \infty $ 
\[
\frac{P_{0,N}(m+2)}{P_{0,N}(m)}\sim \exp (-\mu _{1}+\log 12)(1+\frac{1}{m}+O(%
\frac{1}{m^{2}})) 
\]
and thus as $m\rightarrow \infty $, for example for even length, $%
\lim_{N}P_{0,N}(2m)\sim _{m\rightarrow \in }Cm\exp (m(-\mu _{1}+\log 12))$.
At the same time the second factor in (\ref{main})\ is less than $1$. Thus
from 
\[
Z_{0,N}=\exp (-\mu _{1}\frac{3N}{2})\Theta _{0,N}=\exp (-\mu _{1}\frac{3N}{2}%
)\sum_{m}\exp (-\mu _{1}\frac{m}{2})C_{0}(N,m) 
\]
we get $F=-\frac{3}{2}\mu _{1}+c,c=3\sqrt{\frac{3}{2}}$, as for fixed $m$ we
have 
\[
C_{0}(N,m)\sim \phi (m)N^{-\frac{5}{2}}c^{N} 
\]
From (\ref{main}) we get this formula also in the supercritical case, if we
put $m=\beta N,0<\beta <1$ In the critical point both expressions coincide,
but the free energy is not differentiable at this point.

To prove the second assertion of theorem 2 put $12\exp (-\mu _{1})=1+r$ and
estimate separately all 3 factors in (\ref{main}). We get, that there exists 
$0<\delta \ll \varepsilon \ll 1$ such that for any $m\leq \delta N$ 
\[
\frac{P_{0,N}(m)}{P_{0,N}(\varepsilon N)}<(1+\frac{r}{2})^{-\frac{%
\varepsilon }{2}N} 
\]
The result follows from this.

In the critical case the result follows similarly from from the estimates
(for even $m=\beta \sqrt{N}$)

\[
\frac{P_{0,N}(m+2)}{P_{0,N}(2)}\sim \prod_{k=1}^{\frac{m}{2}}(1+\frac{1}{2k}%
)(1-\frac{4}{3}\frac{k}{N})\sim C\sqrt{m}\exp (-\frac{1}{3}\frac{m^{2}}{N}) 
\]
Then for any $0<\alpha <\beta <\infty $ 
\[
\lim (\frac{P_{0,N}(m(N)<\varepsilon \sqrt{N})}{P_{0,N}(\beta \sqrt{N}%
<m(N)<\beta \sqrt{N})}+\frac{P_{0,N}(m(N)>\varepsilon ^{-1}\sqrt{N})}{%
P_{0,N}(\beta \sqrt{N}<m(N)<\beta \sqrt{N})})=0 
\]
as $\varepsilon \rightarrow 0$.

Let us remove now the coordinate system from the boundary. That is we do not
assume that homeomorphisms respect the origin (the specified edge) on the
boundary. The free energy remains the same. Only in the critical point the
distribution of the length changes - stronger fluctuations appear.

\begin{theorem}
In the critical point without coordinate system the boundary length is of
order $N^{\alpha }$ for any $0<\alpha <\frac{1}{2}$. The exact statement is
that the distribution of $\frac{\log m_{N}}{\log \sqrt{N}}$ converges to the
uniform distribution on the unit interval, that is $P_{0,N}(\frac{\alpha }{2}%
\leq \frac{\log m_{N}}{\log \sqrt{N}}\leq \frac{\beta }{2})\rightarrow \beta
-\alpha $ for all $0\leq \alpha <\beta \leq 1$.
\end{theorem}

Proof. We need the following lemma

\begin{lemma}
$C(N,m)\sim m^{-1}C_{0}(N,m)$ as $N,m\rightarrow \infty $.
\end{lemma}

Proof of the lemma. Let us enumerate the edges of the boundary $1,2,...,m$
in the cyclic order, starting from the root edge. Automorphis $\phi $ of the
triangulation of the disk is uniquely defined if an edge $j=\phi (1)$ is
given, to which the edge $1$ is mapped. In fact, then all adjacent to the
edge $1$ triangles are mapped by $\phi $ to the triangles adjacent to $j$,
and so on by connectedness.

Consider the strip of width $1$, adjacent to the boundary, that is the set
of triangles of \ three types: those which have a common edge with the
boundary (type $1$), two common adjacent edges (type $2$) and those, having
with the boundary only a common vertex (type $0$). Thus the strip can be
identified with the word $\alpha =x_{1}...x_{n},n>m$, where $x_{i}=0,1,2$.
Consider the set $W(m,n_{0},n_{1},n_{2})$ of words with given $m$ and
numbers $n_{i}$ of $i=0,1,2$. \ An automorphism of a disk gives a cyclic
automorphism of the word $\alpha $. The sets $W(m,n_{0},n_{1},n_{2})$ are
invariant however. Note that 
\[
m=n_{1}+2n_{2} 
\]
and the length of other boundary of the strip equals $m^{\prime }=n_{0}$.
Thus, if there is no nontrivial automorphism of the word, then there are no
automorphisms of the whole triangulation of the disk. It is easy to see,
that for a given $n_{0}$ and as $m\rightarrow \infty $ the set of words from 
$\cup _{n_{1},n_{2}}W(m,n_{0},n_{1},n_{2})$, having nontrivial cyclic
automorphisms, is small compared with the number of words $\left| \cup
_{n_{1},n_{2}}W(m,n_{0},n_{1},n_{2})\right| $..

This gives in $f(N,m)$ the factor $1-\frac{1}{m}$ instead of $1+\frac{1}{m}$
in the previous case. Similar calculations show that asymptotically the
distribution coincides with the family $\nu _{N}$ of distributions on the
set $\left\{ 1,...,\sqrt{N}\right\} $\ of $\sqrt{N}$\ elements 
\[
\nu _{N}(i)=Z_{\sqrt{N}}i^{-1},Z_{\sqrt{N}}=\sum_{i=1}^{\sqrt{N}}i^{-1} 
\]
Then it is easy to see that for $0\leq \alpha \leq \beta \leq 1$ 
\[
\nu _{N}(\frac{\alpha }{2}\leq \frac{\log i}{\log \sqrt{N}}\leq \frac{\beta 
}{2})=\nu _{N}(N^{\frac{\alpha }{2}}\leq i\leq N^{\frac{\beta }{2}%
})\rightarrow \beta -\alpha 
\]

\subsubsection{TFT}

Introducing the coordinate system on the boundary resembles transition from
topological field theory to conformal field theory. Topological field theory
(TFT) usually is introduced axiomatically using the category theory \cite
{Ati}, heuristically (using path integrals from quantum field theory) or
algebraically (as Frobenius algebras). to give sense to the path integral
the idea of discretization, see for example \cite{wata, ishka, baez2}, seems
quite natural. In the last chapter of \cite{amdujo} TFT is constructed,
which satisfies all axioms already for finite triangulations. However, it is
possible that for some cases the axioms can be satisfied only in the limit $%
N\rightarrow \infty $. Moreover, such limit is interesting in itself, and
little is known about it. Here we consider this problem for two boundaries
(that is $k=2$), zero genus and large $\mu _{1}$ in the previous model. It
is interesting that we get a nontrivial joint distribution of the lengths of
the boundaries.

\begin{theorem}
For $k=2$, sufficiently large $\mu _{1}$ and fixed $m_{1},m_{2}$ there exist
correlation functions $P_{0,2}(m_{1},m_{2})=\lim_{N}P_{N,0,2}(m_{1},m_{2})$.
\end{theorem}

We give a complete proof, similar method seems to work for any $k$ and $\rho 
$. The following recurrent relations for $C_{0}(N,m,m_{2}),m=m_{1}$ hold

\[
C_{0}(N,m,m_{2})=C_{0}(N-1,m+1,m_{2})+C_{0}(N-1,m+m_{2}+1)+ 
\]
\[
+\sum_{N_{1}+N_{2}=N-1}%
\sum_{k_{1}+k_{2}=m+1}C_{0}(N_{1},k_{1},m_{2})C_{0}(N_{2},k_{2}) 
\]
They can be obtained as in Tutte's method \cite{tutte1} by deletion of the
rooted edge. See Figure ???

\begin{figure}[tbp]
\includegraphics[0,0][40,100]{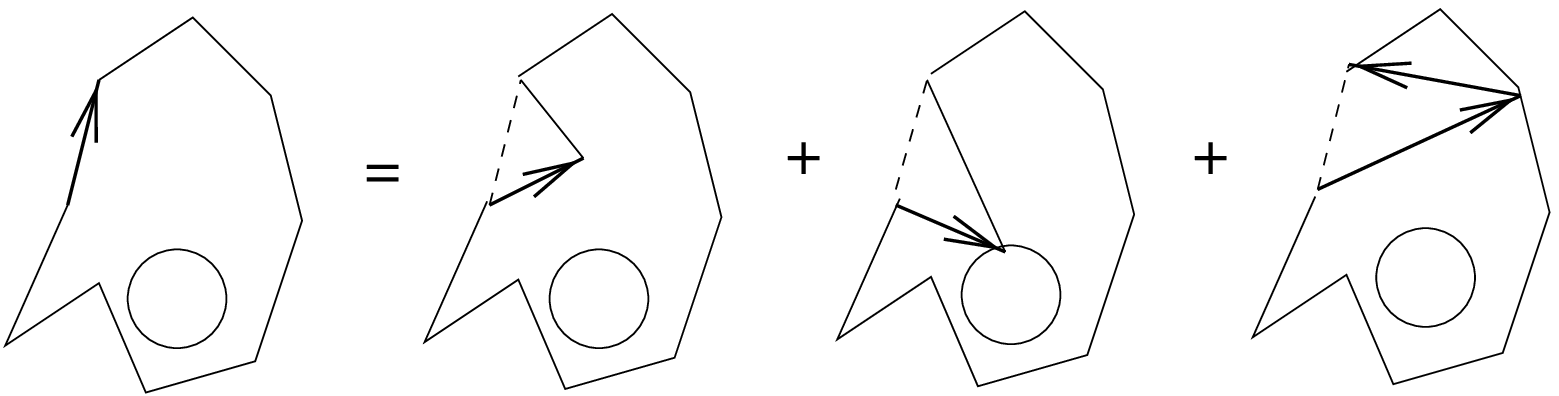}
\caption{{}}
\end{figure}
After deletion three cases are possible which correspond to three terms in
the right hand side of the recurrent relation: graph has still two holes but
the first one (where the root is) the length increases but the number of
triangles decreases, two holes become one, the graph becomes not connected.
The boundary conditions are as follows. For given $N$ only finite number of
terms $C_{0}(N,m,m_{2})\neq 0$, it is convenient to consider $%
C_{0}(N,m,m_{2})$\ known for $N<N_{0}$, their explicit expression does not
interest us. Moreover, $C_{0}(N,m,m_{2})=0$, if either $m<2$ or $m_{2}<2$.

\begin{lemma}
The asymptotics as $N\rightarrow \infty $ and fixed $m,m_{1}$%
\begin{equation}
C_{0}(N,m,m_{2})\sim \phi (m,m_{2})N^{-\frac{5}{2}+1}c^{N}  \label{as2}
\end{equation}
Moreover $\phi (m,3)=\phi (m)$.
\end{lemma}

Proof. It is well-known that 
\begin{equation}
C_{0}(N,m)\sim \phi (m)N^{-\frac{5}{2}}c^{N}  \label{as4}
\end{equation}
Find first $C_{0}(N,m,2),C_{0}(N,m,3),C_{0}(N,2,m_{2}),C_{0}(N,3,m_{2})$.
Let first $m_{2}=3$. To get a triangulation from $T(0,N;m,3)$\ take a
triangulation from $T(0,N+1;m)$ and choose one of $N$ triangles so that it
did not touch the boundary, this can be done in $N-O(m)$ ways. It follows
that $C_{0}(N,m,3)\sim C_{0}(N+1,m)N$, and thus $\phi (m,3)=c\phi (m)$.
Forgetting the root on the boundary $m$ and putting it on the boundary of
length $3$, we have from triviality of the automorphism group fro almost all
graphs that $\phi (3,m)=\frac{3c}{m}\phi (m)$. To get a triangulation from $%
T(0,N;m,2)$\ take a triangulation from $T(0,N;m)$, choose one of its $N$
triangles so that it did not touch the boundary, take one of its edges and
double it. That is why $\phi (m,2)=\frac{3}{2}\phi (m)$ and similarly $\phi
(2,m)=\frac{3}{m}\phi (m)$.

Rewrite the recurrent relations as follows 
\[
C_{0}(N-1,m+1,m_{2})=C_{0}(N,m,m_{2})-\sum_{N_{1}+N_{2}=N-1}%
\sum_{k_{1}+k_{2}=m+1}C_{0}(N_{1},k_{1},m_{2})C_{0}(N_{2},k_{2})+ 
\]
\[
-C_{0}(N-1,m+m_{2}+1) 
\]
As far as we know $C_{0}(N,m,2),C_{0}(N,m,3)$ and $%
C_{0}(N,2,m_{2}),C_{0}(N,3,m_{2})$, the result follows by induction in $m$
for fixed $m_{2}$ and vice versa. In fact, multiply the latter relation on $%
c^{-N+1}(N-1)^{\frac{5}{2}-1}$. Then by inductive assumption the right hand
side tends to 
\[
c\phi (m,m_{2})-2\sum_{k_{1}+k_{2}=m+1}\phi (k_{1},m_{2})\sum_{n=1}^{\infty
}C_{0}(n,k_{2})c^{-n} 
\]
Thus, the limit of the left hand side $c^{N-1}(N-1)^{\frac{5}{2}%
-1}C_{0}(N-1,m+1,m_{2})$ is finite, denote it $\phi (m+1,m_{2})$.

Come back to the proof of the theorem. Substituting (\ref{as4}) to the
recurrent formula for $C_{0}(N,m)$ we get recurrent formula for $\phi (m)$%
\[
\phi (m+1)=c\phi (m)-2\sum_{k_{1}+k_{2}=m+1}\phi
(k_{1})\sum_{n}C_{0}(n,k_{2})c^{-n} 
\]
In particular 
\[
\phi (3)=c\phi (2),\phi (4)=c\phi (3)-2\phi (2)\sum_{n}C_{0}(n,2)c^{-n} 
\]
Now the asymptotics for $\phi (m,m_{2})$ can be obtained from the relation 
\[
\phi (m+1,m_{2})=c\phi (m,m_{2})-2\sum_{k_{1}+k_{2}=m+1}\phi
(k_{1},m_{2})\sum_{N_{2}=1}^{\infty }C_{0}(N_{2},k_{2})c^{-N_{2}} 
\]
It follows $\phi (m,m_{2})<\phi (2,m_{2})c^{m}$. But $\phi (2,m_{2})=\frac{2%
}{m_{2}}\phi (m_{2},2)<\phi (2,2)c^{m_{2}}$. Theorem is proved.

\subsection{Nonuniqueness for a low entropy case}

Let $U_{2}$ be the graph isomorphic to the $2$-neighborhood of the origin of
the lattice $Z^{2}$. Then among graphs belonging to $\mathcal{G}(U_{2})$
there is the lattice $Z^{2}$ itself and its factor groups $%
Z(k_{1},0),Z(k_{1},k_{2})$\ with respect to subgroups $\left\{
(nk_{1},0)\right\} $ and $\left\{ (n_{1}k_{1},n_{2}k_{2})\right\}
,n,n_{1},n_{2}\in Z$, that is cylinders and tori correspondingly, where $%
k_{i}\geq 4$. Note that there are other graphs in $\mathcal{G}(U_{2})$, for
example twisted cylinders, which can be obtained from the strip $Z\times
\left\{ 0,1,2,...,k\right\} \subset Z^{2}$ by identifying the points $(n,0)$
and $(n+j,k)$ for any $n$. If there is no spin then the entropy has a
subexponential growth.\ 

Consider a Gibbs families with $S=\left\{ -1,1\right\} $ and potential $\Phi 
$, equal to the sum of $\Phi _{U_{d,2}}$ and the Ising potential $\Phi
_{Is}=\sum_{<i,j>}\sigma _{i}\sigma _{j}$ of nearest neighbor interaction.

\begin{theorem}
For any $\beta $\ there exists infinite number of pure Gibbs families with
potential $\Phi $, that are Gibbs fields on fixed graphs (cylinders).
\end{theorem}

Proof. We shall construct for sufficiently large $N$ the finite Gibbs family
with boundary conditions, defined by the graph $\gamma _{N}$, which is
constructed as follows. Consider the subgraph $\alpha _{n}$ of the graph $%
Z^{2}$ - the strip, consisting of points $(i,j),i\in Z;j=-n,...,n,n\geq 2$,
with root $(0,0)$. The points $(i,-n)$ and $(i,n)$ identified. Define the
distance of the point $(i,n)$ to $(0,0)$ as $\left| i\right| +\left|
n\right| $. Put $\gamma _{N}=\gamma (\alpha _{n},(0,0);N+1,N+3)$. \ If $n\ll
N$, then $\gamma _{N}$\ consists of two isomorphic connected components. We
will use lemma \ref{lem1}, assuming that $\nu _{N}$ is the unit measure on
the spin graph $\gamma _{N}$, taking all spins on $\gamma _{N}$ equal to\ $1$%
. We shall prove that finite Gibbs family on $\mathcal{G}_{N}^{(0)}$ is the
Gibbs (Ising) measure on the graph $\gamma (\alpha _{n},(0,0);0,N)$.

To prove this it is convenient to start from the case when $S$ is trivial,
using only part $\Phi _{U_{d,2}}$ of the potential $\Phi $. Prove that the
finite Gibbs family with the defined boundary conditions is a unit measure
on $\gamma (\alpha _{n},(0,0);0,N)$, that is the random graph $G$ coincides
a.s. with the graph $\gamma (\alpha _{n},(0,0);0,N)$. We shall use the
method of ''analitical'' continuation, consisting in the inductive proof of
the coincidence of the slice $\gamma (N)=\gamma (G,0;N,N),\gamma (N-1)$ etc.
of the random graph $G$ coincide with the correponding annuli of the graph $%
\alpha _{n}$. We use double induction. Construct first the annulus $\gamma
(N)$ and prove that it is unique and coincides with the corresponding
annulus $\gamma (\alpha _{n},(0,0);N,N)$\ of the graph $\alpha _{n}$.

Consider the point $(N+2,0)$ in $\gamma _{N}$\ and its $2$-neighborhood in $%
\gamma _{N}$. This neighborhood lacks one edge $l_{0}$ from the point $%
(N+1,0)$ with the new vertex $v_{0}$, which we denote $(N,0)$. The latter
vertex should be on the distance $N$ from the origin as all point on
distances $N+1,N+2;N+3$ are already fixed.

Consider now the point $(N+1,1)$. From the geometry of its neighborhood in $%
\gamma (\alpha _{n},(0,0);N+1,N+3)$ it follows that from the point $(N,1)\in
\gamma (\alpha _{n};N+1,N+3)$ there should be two new edges $l_{1},l_{2}$
connecting it with the slice (annulus) $N$. Moreover, one of these edges
should be connected with the vertex $v_{0}$.

Furthe by induction we construct all slice $\gamma (N)$ (and only this
slice), starting with points with positive first coordinate and then
symmetric ones. All points of the slice $\gamma (N+2)$ will get necessary
neighborhoods and one can exclude them from further considerations.

Further on by induction in slices we construct slices $\gamma (N-1),\gamma
(N-2)$, ... , using the neighborhoods of the already constructed slices $%
N+1,-N-1,N,-N,...$. We continue this induction until both connected
components become one. Let us prove that this event will occur on the line $%
y=\pm n$ on slice $n$. We should exclude two other possibilities: the first
one - that this will occur on another line, second o,e - that it will occur
on the line $y=\pm n$, but on a slice different from $n$. The second
possibility can be excluded with the following arguments. After murging of
the two connected components on the line $y=\pm n$ we will have a hol which
should be filled in for exactly $n$ steps of the induction. That is why the
murging should occur on the slice $n$. The first possibility is excluded as
follows. Assume that the murging occured on the line $y=k$. Then the line $%
y=k+1$ on this step will have on the same inductive step an obstruction to
filling it in.

Thus, we see that for any choice of the spins on $\gamma _{N}$ the Gibbs
family has a support on the unique graph, thus thois Gibbs family is a Gibbs
field. Theorem is proved as $n$ was arbitrary.

\subsection{Gibbs characterization of structures}

\subsubsection{Uniqueness for high entropy case}

Countable trees cannot be characterized locally, that is they cannot be
selected from the class of all graphs using some restrictions on subgraphs
with diameter less than some constant. It is interesting however, that a
local characterization (understood probabilistically) can be given using
Gibbs families with local potentials. More exactly, one can define Gibbs
families with potential of diameter $2$, which have support on the set of
countable tress. We consider here the case of $p$-regular trees. Moreover,
this will give us an example when the partition function has factorial
growth but the corresponding Gibbs family is unique.

Consider the set $\mathcal{A}_{N,p}$ of all $p$-regular graphs with $N$
vertices, that is graphs where all vertices have degree $p\geq 3$. Consider
first the finite Gibbs family on $\mathcal{A}_{N}$ with the following
superpotential $\Phi :$ $\ \Phi =0$ on graphs of radius $1$, if $0$ has
degree $p,$ and $\Phi =\infty $ for all other graphs of diameter $1$. One
can say also that we consider the Gibbs family $\mu _{N}$ on $\mathcal{A}%
_{N,p}$ with the potential $\Phi \equiv 0$.

\begin{theorem}
\label{p-reg}$\lim p^{N}(\Gamma _{k})=p(\Gamma _{k})=1$ , if $\Gamma _{k}$
is any $p$-regular, where each path from $0$ to a final vertex has length $k$%
, and $0$ for other $\Gamma _{k}$.
\end{theorem}

Proof. Note that all graphs from $\mathcal{A}_{N,p}$ have equal probability
with respect to the measure $\mu _{N}$, that reduces the problem to the
classical theory of random graphs, see \cite{bol}. The combinatorial method
to prove this theorem consists of several steps.

\begin{enumerate}
\item  Let us call a graph enumerated, if its vertices are enumerated. Let $%
L_{N}(p)$ be the number of enumerated $p$-regular graphs with $N$ vertices.
This number is equal to, see theorem II.16 in \cite{bol}, 
\[
L_{N}(p)\sim C(p)\frac{(pN-1)(pN-3)...}{(p!)^{N}} 
\]
where $C(p)$ is some constant, which is known explicitely but we do not need
it. Similarly for fixed $p,k,d_{1},...,d_{k}$ let $L_{N}(p;d_{1},...,d_{k})$
be the number of enumerated graphs with where some $N-k$ vertices have
degree $p$ and other $N-k$ vertices have degrees $d_{1},...,d_{k}$
correspondingly. It is equal to, see also theorem II.16 in \cite{bol}, 
\[
L_{N}(p;d_{1},...,d_{k})\sim C(p,k;d_{1},...,d_{k})\frac{(2m-1)(2m-3)...}{%
(p!)^{N}} 
\]
where $C(p,k;d_{1},...,d_{k})$ is a constant and $2m=p(N-k)+d_{1}+...+d_{k}$.

\item  Prove that the probability that the neighborhood $O_{1}(v)$ of some
vertex (for example, of the vertex $1$) does not have cycles, tends to zero
as $N\rightarrow \infty $. The number of graphs such that $O_{1}(v)$ does
not contain cycles is equal to $L_{N-1}(p;d_{1},...,d_{p})(N-1)...(N-p)$
with $d_{1}=...=d_{p}=p-1$. At the same time the number of graphs such that $%
O_{1}(v)$ contains one cycle, that is there is an additional edge, say
between edges $1$ and $2$, is equal to $%
L_{N-1}(p;e_{1},...,e_{p})(N-1)...(N-p)$, where $%
d_{1}=d_{2}=p-2,d_{3}=...=d_{p}=p-1$. It is easy to see that 
\[
\frac{L_{N-1}(p;e_{1},...,e_{p})}{L_{N-1}(p;d_{1},...,d_{p})}\rightarrow 0 
\]
as $N\rightarrow \infty $.

\item  It follows that the mean number of vertices $v$ with neighborhoods $%
O_{1}(v)$ without cycles is of order $o(N)$. But as almost all $p$-regular
graphs do not have nontrivial automorphisms (theorem IX.7 \cite{bol}), then
the same result holds for non enumerated graphs.

\item  Neighborhoods or larger radius are considered similarly.
\end{enumerate}

Let us do an important remark: it is not true that for a given $N$ the
probability for the graph in $\mathcal{A}_{N,p}$ to be a tree, is close to $%
1 $. Moreover, this probability tends to $0$ as $N\rightarrow \infty $.
Nontrivial topology (cycles) appears on a larger scale: typical length of
cycles tends to $\infty $ as $N\rightarrow \infty $, see below.

\begin{remark}
If the set $S$ is finite, but the interaction is trivial, gives an
independent fields on the $p$-regular tree. The question is what will be if
there an additional nontrivial interaction between spins.
\end{remark}

\subsubsection{Gibbs varieties of groups}

Beside the trees, Gibbs families allow to select other mathematical objects,
which do not have a local characterization. WE consider the cases when after
the limit a.s. we get an algebraic object - a group.

Consider the class $\mathcal{H}_{N,2p}$ $2p-$regular graphs with a local
structure, where all edge-ends of each vertex are enumerated by symbols $%
a_{1},...a_{p},a_{1}^{-1},...a_{p}^{-1}$. Moreover, if to one end of an edge
there corresponds say $a_{i}$, then to the other end there corresponds $%
a_{i}^{-1}$, and vice-versa. This class can be selected from the class $%
\mathcal{A}_{N}$ with the local potential $\Phi $ on edges, equal $0$, if
the edge-ends have symbols $a_{i},a_{i}^{-1}$, and $\infty $ otherwise, and
also by the superlocal potential $\Phi _{1}$ on neighborhoods of radius $1$,
equal $0$ only if the vertex has degree $2p$ and all its edge-ends are
enumerated by different symbols.

In particular, Cayley diagrams of countable groups, see \cite{makaso},
belong to this class of graphs with a local structure. The graph of the
group with finite number of generators $a_{1},...a_{p}$ and finite number of
defining relations $\alpha _{1}=...=\alpha _{n}=1$, where $\alpha
_{1},...,\alpha _{n}$ are some words, is selected by the following nonlocal
condition: any cycle in the graph is generated by $\alpha _{1},...,\alpha
_{n}$, and vice versa (note that to each path on the graph corresponds a
word, that is an elemnet of the group). The free noncommutative group
corresponds to the case $n=0$, that is when there are npo defining relations.

\begin{theorem}
Limiting Gibbs family on $\mathcal{A}_{\infty }^{(0)}$ with potential $\Phi
+\Phi _{1}$\ has the support on the tree corresponding to the free group
with $n$\ generators.
\end{theorem}

We do not give proof here, but note that it differs from the previous proof
for $p$-regular graphs only in that the set edge-ends is subdivided on $2p$
subsets $A(a_{i}),A(a_{i}^{-1})$, and the freedom of coupling edge-ends is
restricted only by that edge-ends, for example $A(a_{i})$, can be coupled
only with edge-ends from $A(a_{i}^{-1})$.

This result possibly allows wide generalizations. Consider the set $\mathcal{%
H}_{N,2p}(\alpha _{1},...,\alpha _{n})$ of graphs, corresponding to groups
from a manifold of groups, defined by relations $\alpha _{1}=...=\alpha
_{n}=1$ (and possibly others as well). Then the limiting Gibbs families has
its support on the free with respect to this manifold groups (that is having
only reltions $\alpha _{1}=...=\alpha _{n}=1$), as the creation of
additional short cycles have small probability. Such results can play role
in physics explaining how nonlocal objects can appear with great probability
from local conditions.

\subsection{Scales}

Here we shortly say about an important generalization, that allows to go out
of the framework of the thermodynamic limit which only leads to limiting
Gibbs families.

Let Gibbs families $\mu _{N}$ on the sets $\mathcal{B}_{N}$ of finite graphs
with a local structure be given, where $N$ is some parameter as the number
of vertices, edges, radius etc. They induce measures $\nu _{N}$ on the
corresponding subsets $\mathcal{G}_{N}$ of finite graphs. Let a
nondecreasing function $f:Z_{+}\rightarrow R_{+}$ be given. Define the
macrodimension of a random (with respect to the sequence $\nu _{N}$) finite
graph on the scale $f(N)$. Let $O_{f(N)}(v)$ be the neighborhood of $v$ of
radius $f(N)$. Put 
\[
D_{N}(f)=<\frac{1}{\left| V(G)\right| }\sum_{v\in V(G)}\frac{\log O_{f(N)}(v)%
}{\log f(N)}>_{\nu _{N}} 
\]
If there exists the limit $\lim_{N\rightarrow \infty }D_{N}(f)=D(f)$, then $%
D(f)$ is called the macrodimension on the scale $f$.

One can define the corresponding invariant for limiting Gibbs families,
using the limit $N\rightarrow \infty $. See, for example, the definition of
the macrodimension of a countable complex in \cite{m6}. Such definition
however is only a particular case, it corresponds to the minimal scale $f$,
where $f(N)$ tends to infinity with $N$ but as slower than any othe fixed
sclae.

Similarly one can define other topological characteristics on different
scales. Let $\mathcal{B}_{N}=\mathcal{A}_{N}$, that is $V(G)=N$. For
example, let $h(O_{d}(v))$ be the number of independent cycles in the $d$%
-neighborhood of the vertex $v$. The exponent for the number of independent
cycles on the scale $f$ is defined as follows, if the following limit
exists, 
\[
b(f)=\lim_{N\rightarrow \infty }\frac{1}{N}<\sum_{v}\frac{h(O_{f(N)}(v))}{%
\left| O_{f(N)}(v)\right| }> 
\]

\begin{proposition}
Under the conditions of theorem \ref{p-reg} $b(f)=0$ for any scale $f(N)\leq
(1-\varepsilon )\log _{p-1}N$, where $\varepsilon >0$ is arbitrary, and is
equal to $\frac{p}{2}-1$ for any scale $f(N)\geq (1+\varepsilon )\log
_{p-1}N $.
\end{proposition}

Proof. Consider the neighborhood $O_{(1-\varepsilon )\log _{p-1}N}(v)$ of
some vertex $v$. In this neighborhood there is not more than $%
(p-1)^{(1-\varepsilon )\log _{p-1}N}=N^{1-\varepsilon }$ vertices. An edge
outgoing from these vertices is incident to another vertex of this
neighborhood with probability not greater than $N^{-\varepsilon }$. Then the
number of such edges does not exceed $N^{-\varepsilon }pN^{1-\varepsilon
}=pN^{1-2\varepsilon }$. Any such edge cannot create more than one new
cycle. The first statement follows from this.

To prove the second statement denote $D(N)$ the diameter of the random
graph. It is known \cite{bol}, that fro any $\varepsilon >0$ 
\[
P_{N}(1-\varepsilon <\frac{D(N)}{\log _{p-1}N}<1+\varepsilon )\rightarrow
_{N\rightarrow \infty }0 
\]
In all graph the number of independent cycles is $(\frac{p}{2}-1)N-1$, that
follows from Euler formula $V-L+M=1$, where $M$ is the number of independent
cycles. Then $M=L-V=\frac{pN}{2}-N$.

\subsection{On discrete quantum gravity}

Here it is natural to say some words about main directions of research in
discrete quantum gravity. The question ''What is quantum gravity?'' has no
exact formulation. There are many attempts to formulate and solve this
problem (however, an extreme optimism in some physical papers always changed
with more moderate attitude). As we will see now, most approaches has much
in common with the scheme developed in this paper. However, the main
unsolved difficulty, not mentioning the necessity of fundamental ideas,
consists in concretization of the model, following numerous physical
requirements, which are unclear as well.

Physicists claim that the so called standard model satisfies practical
needs, thus describing all interactions (electromagnetic, weak, strong),
except gravity. Although the standard model does not still exist in
space-time dimension 4 from the point of view of constructive quantum field
theory (that is from the standard point of view on mathematical rigor, it
seems to admit at least exactly formulated definitions and
assertions-hypothesis. They can be given either generalizing Wightman axioms
or as scaling limits of lattice approximations.\ However, in all attempts to
construct unified theory, including gravity, even exact hypothesis are
lacking. What exists now are some pieces of ''future theory'', based on
different physical, methodological and even philosophical principles. I will
try to give some classification of them, staying in a mathemical background.

Note first, that no ways are known now for direct generalizations of
Wightman axioms or, in euclidean approach, Osterwalder-Schrader axioms
(although, axioms of $S$-matrix type can be generalized in different
directions - topological field theory, diagrams of string theory). Main
underlying conceipts of these axioms are based on the classical and fixed
space-time. For example, locality uses Minkowski metrics, and unitarity -
the mere existence of classical time. If, in the framework of Wightman
axioms one adds the metrics as a new field, then it will not be (as the
experience in constructive field theory shows) continuous function, that can
change topology locally, making the conceipt of locality unclear. One says
then that metrics quantization leads to space-time quantization.

Lattice approximation is a more convenient object for generalization. In
fact, all existing theories allow discrete approximation. For example,
euclidean field theory (restricting to boson fields for simplicity)
considers the lattice $Z_{\varepsilon }^{d}$ with step $\varepsilon $,
volume $\Lambda =\left[ -N,N\right] ^{d}\subset R^{d}$, fields $\phi (x)\in
R^{m},x\in Z_{\varepsilon }^{d},$ and the partition function 
\[
Z_{\Lambda ,\varepsilon }=\left[ \prod_{x\in \Lambda \cap Z_{\varepsilon
}^{d}}\int_{R^{m}}d\phi (x)\right] \exp (-\sum_{<x,x^{\prime }>}\Phi (\phi
(x),\phi (x^{\prime });\varepsilon )) 
\]
where $x,x^{\prime }$is any pair of nearest neighbors in $\Lambda \cap
Z_{\varepsilon }^{d}$, $\Phi $ is a real function on $R^{m}\times R^{m}$,
depending also on a parameter $\varepsilon $. Main mathematical problem is
the existence of some scaling limits as $\varepsilon \rightarrow 0$.

In this paper we can only touch different directions for constructing
discrete models. We show that they are Gibbs families by explicitely
indicating the classes of graphs with a local structure and the potential.
Wide physical reviews of different directions in quantum gravity see in\ 
\cite{gib, wal, rov2}.

It is important to note that most theories listed below try to quantize
metrics but not topology, as the considered class of graphs is related to
one fixed manifold or with some class of manifolds of rather restricted
topology. Thus, a part of the theory seems to be the same for all existing
discrete theories, and the theory of Gibbs families seems to be necessary.

\paragraph{Regge theory}

Chronologically this is the earliest (see\ \cite{Fro}) discrete alternative
to classical general relativity. As compared to more recent dynamical
triangulation model, where the lengths of edges are considered to be equal,
they are arbitrary in Regge theory. Let us fix a triangulation $T$ of a
smooth manifold $M$ of dimension $d$ and let $S_{k}(T),k=0,...,d,$ be the
set of all $k$-simplices of the triangulation $T$. Let $l_{i}$ be the length
of edge $i\in S_{1}(T)$ and $\mathcal{L}(T)=\left\{ l_{i},i\in
S_{1}(T)\right\} $ be the set of vectors of lengths, such that on each
simplex they define a positive definite euclidean metrics. Thus in euclidean
approach $l_{i}$ are random variables, satisfying some inequalities, that
makes the free measure more complicated. The partition function is defined
to be 
\[
Z_{N}=\sum_{T:\left| T\right| =N}\exp (-S(T))d\Lambda (T),S(T)=\lambda
\sum_{B\in S_{d-2}(T)}v(B)\varepsilon (B)+\mu \sum_{\Delta \in
S_{d}(T)}v(\Delta )
\]
where $d\Lambda (T)$ is the restriction of the measure $\prod_{i\in
S_{1}(T)}dl_{i}$ on $\mathcal{L}(T)$, $v(.)$ is the volume, $\varepsilon (B)$
is the defect angle of the $(d-2)$-dimensional face $B$, which we will not
define here (see \cite{Fro}). They can be expressed in terms of $l_{i}$.

\paragraph{Dynamical triangulations}

This is possibly the domain most rich in results. It appeared in the string
theory community. Note that, out of all directions in quantum gravity, only
string theory includes the standard model and tries to get numbers from the
theory. \ Although string theory follows tightly the formal techniques of
quantum field theory, there is no question about axioms. String theory has
more calculational goals (to calculate string diagrams) than analysis of
fundamentals of our knowledge about space-time.

Discrete string in euclidean approach is a Gibbs family on some subsets $%
\mathcal{A}_{N}$ of the set $\cup _{\rho ,k}\mathcal{A}_{N}(\rho ,k)$ of
triangulations $T$ of surfaces of genus $\rho $ with $k$ holes. On each
triangle $i$ there is a spin $\sigma _{i}$ taking values in $R^{d}$. Often $%
\mathcal{A}_{N}$ is taken to be $\mathcal{A}_{N}=\mathcal{A}_{N}(0,0)$ or $%
\cup _{\rho }\mathcal{A}_{N}(\rho ,0)$. The canonical partition function is 
\[
Z_{N}=\sum_{T}\int_{\mathcal{A}_{N}}\exp (-\lambda \rho
(T)-\sum_{<i,j>}(\sigma _{i}-\sigma _{j})^{2})\prod_{i\in T}d\sigma _{i}
\]
Nondiscrete approaches to string theory has more popularity. For example,
the spectrum of the free boson string can be found explicitely via the
hamiltonian approach, but this is not yet done (and seems to be difficult)
in the dynamical triangulation approach. This enigmatic fact shows that
relations between different approaches to physically the same object can be
very subtle.

\paragraph{Matrix theories}

Matrix models are can be used for counting maps (and more complicated
objects) on surfaces. Smooth map is a triple $(S,G,\phi )$, where $S$ is a
smooth compact oriented surface, $G$ is a connected graph (one-dimensional
complex), $\phi :G\rightarrow S$ is an embedding such that the images of the
edges of $G$ are smooth arcs, and the complement $S\backslash \phi (G)$ is a
union of domains homeomorphic to disks. A map (or combinatorial map) is an
equivalence class of smooth maps. Two maps $(S_{1},G_{1},\phi
_{1}),(S_{2},G_{2},\phi _{2})$ are called equivalent if there exists a
homeomorphism $f:S_{1}\rightarrow S_{2}$ such that mappings $f:\phi
_{1}(V(G_{1}))\rightarrow \phi _{2}(V(G_{2}))$ and $f:\phi
_{1}(L(G_{1}))\rightarrow \phi _{2}(L(G_{2}))$\ are one-to-one.

One of the central models of matrix theory is the following probability
distribution $\mu $ on the set of selfadjoint $n\times n$-matrices $\phi
=(\phi _{ij})$ with the density 
\[
\frac{d\mu }{d\nu }=Z^{-1}\exp (-tr(\frac{\phi ^{2}}{2h})-tr(V))
\]
where $V=\sum a_{k}\phi ^{k}$ is a polynomial of $\phi $, bounded from
below, $\nu $ is the Lebesgue measure on the real $n^{2}$-diumensional space
of vectors $(\phi _{ii},\func{Re}\phi _{ij},\func{Im}\phi _{ij},i<j)$. If $%
V=0$, then the measure $\mu =\mu _{0}$ is gaussian with covariances $%
\left\langle \phi _{ij},\phi _{kl}^{\ast }\right\rangle =\left\langle \phi
_{ij},\phi _{lk}\right\rangle =h\delta _{ik}\delta _{jl}$. The density of $%
\mu $ with respect to $\mu _{0}$ is equal to 
\[
\frac{d\mu }{d\mu _{0}}=Z_{0}^{-1}\exp (-tr(V))
\]
For the existence of $\mu $ it is necessary that the degree $p$ of the
polynomial $V$ were even and the coefficient $a_{p}$ were positive. For this
case there is a deep theory of such models, see \cite{Pas}.

Fundamental connection (originated by Hooft) between matrix models and
counting problems for maps on surfaces is given by the formal series in
semiinvariants or in diagrams (see for example \cite{MalMin1}) 
\[
\log Z=\sum_{k=1}^{\infty }\frac{(-1)^{k}}{k!}<tr(V),...,tr(V)>=\sum_{k=1}^{%
\infty }\frac{(-1)^{k}}{k!}\sum_{D_{k}}I(D_{k})
\]
where $\sum_{D_{k}}$ is the sum over all connected diagrams $D_{k}$ with $k$
vertices and $L=L(D_{k})$ edges. Assume, for example, $V=a\phi ^{4}$. Then
each diagram has enumerated vertices $1,...,k$, has $L=2k$ edges, and
moreover each vertex has enumerated edge-ends (legs) $1,2,3,4$,
corresponding to all factors of the product $\phi _{ij}\phi _{jk}\phi
_{kl}\phi _{li}$. The leg, for example corresponding to $\phi _{ij}$, can be
imagined as a strip (ribbon) (defining thus a ribbon or fat graph), the
sides of the leg-ribbon have indices $i$ and $j$ correspondingly. In a
neighborhood of the vertex the ribbons are placed on the surface as on the
Figure. It is assumed that the coupling of ribbon legs is such that the
coupled sides have the same indices.

\begin{figure}[tbp]
\includegraphics[0,0][40,100]{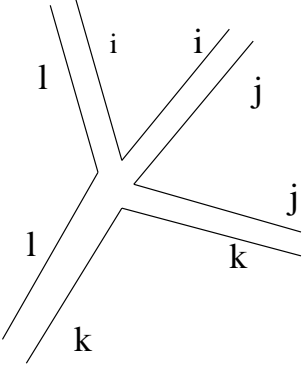}
\caption{{Vertex of a ribbon graph}}
\end{figure}

As any index appears an even number of times in each vertex, then for a
given side $l$ with index $i$, there is a unique closed connected path in
the diagram consisting of sides with index $i$, and passing through $l$.
These paths are called index loops. Summing over indices we will get then
the factor $n^{N}$ where $N=N(D_{k})$ is the number of index loops in the
digram. Finally we get the following formal series 
\[
\sum_{k}\frac{(-a)^{k}}{k!}\sum_{D_{k}}h^{2k}n^{N(D_{k})}=%
\sum_{k}(-4ah^{2})^{k}\sum_{E_{k}}n^{N(E_{k})}=%
\sum_{k,N}(-4ah^{2})^{k}n^{N}M(k,N)
\]
where in the second sum the summation is over all graphs with unordered set
of vertices, and the set of legs of each vertex is only cyclically ordered.
While changing from $D$ to $E$ we introduced the factor $\frac{k!4^{k}}{A(E)}
$, where $A(E)$ is the power of the automorphism group of $E$. We omit $A(E)$%
, assuming that for ''almost all'' (as $k\rightarrow \infty $) graphs $A(E)=1
$.

In the third sum $M(k,N)$ is the number of maps with $k$ vertices and $N$
faces. In fact, it is not difficult to see that for any graph $E$ there
exists a unique (up to combinatorial equivalence) embedding $f(E)$ of this
graph to some oriented compact closed surface $S_{\rho }$ for some genus $%
\rho $ and such that each index loop bounds an open subdomain of $S_{\rho }$
homeomorphic to a disk. This map has $k$ vertices, $2k$ edges and $N$ faces.
By Euler formula $k=N+2\rho -2$ we get the following expansion, putting $%
h=1,a=\frac{b}{n}$ 
\[
\log Z=\sum_{N,\rho }(-4b)^{N+2\rho -2}n^{-2\rho +2}M_{\rho }(N)
\]
where $M_{\rho }(N)$ is the number of maps with $N$ faces and having genus $%
\rho $. It follows that for example for $\rho =0$, the sum of terms
corresponding to maps with $N$ faces and having genus $\rho $ can be
formally obtained as 
\[
\lim_{n\rightarrow \infty }\frac{\log Z}{n^{2}}=\sum_{N}(-4b)^{N-2}M_{0}(N)
\]
In general one can cay that application of the matrix method to counting of
maps is rigorous only in case of triangulations with one (see \cite{HaZa})
or fixed finite number of cells (see \cite{Konts1, Okou1}). It is difficult
to get a nonformal justification of the last limit. Very clearly written
introduction to applications of the matrix method to counting problems is
presented in \cite{Zvo}. Complete but less mathematical presentations one
can find in many physical papers, see \cite{BrItPaZu}. The case of fixed $%
\rho $ (this corresponds to counting string diagrams in perturbation theory)
is well studied only for the dimension $d=0$ of the target space, and for
the following generalization with $Q$ matrices $M_{q}$, where the action is 
\[
Tr(\sum_{q=1}^{Q}V(M_{q})+\sum_{q=1}^{Q-1}M_{q}M_{q+1})
\]
\qquad and for its continuum analog as well, see review \cite{Kaz1}.

One should note that the case of large $\rho $ (that is growing linearly
with $N$), on the physical language this corresponds to the space-time foam
in the sense of Hawking, is open even with the absence of spin, that is for $%
d=0$.

\paragraph{Spin foam}

Spin foam is a graph $\Gamma $ together with the local structure of a
two-dimensional complex (it is assumed that each vertex and each edge belong
at least to one two-dimensional cell) and functions $a(f)$ on the set $%
\Gamma ^{(2)}$ of 2-cells $f$ and $b(l)$ on the set $\Gamma ^{(1)}$ of
edges. Moreover, $a(f)$ defines an irreducible representation of some group
(Lee or quantum) $G$, let $H_{a(f)}$ be the Hilbert space of this
representation. If the edge $l$ is incident to the faces $f_{1},...f_{k}$,
then put $H_{l}=H_{a(f_{1})}\otimes ...\otimes H_{a(f_{k})}$. Let $H_{l}^{0}$
be an invariant space in $H_{l}$. Choose once and for all in the invariant
spaces of tensor products of any irreducible representations some
orthonormal bases. Then $b(l)$ is defined to be an element of the
corresponding basis.

On a class of finite graphs $\mathcal{A}$ (for example, with $N$ vertices)
with such localo structure define the following partition function 
\[
Z_{\mathcal{A}}=\sum_{\Gamma \in \mathcal{A}}Z(\Gamma ),Z(\Gamma
)=\sum_{\left\{ a(.),b(.)\right\} }\prod_{f\in \Gamma ^{(2)}}\dim
a(f)\prod_{v\in \Gamma ^{(0)}}\Phi (O_{1}(v))
\]
where $\Phi $ is some function of the neighborhood of radius 1 of the vertex 
$v$. If $\Phi $ is positive then this defines some Gibbs family.

As far as I know such partition functions were considered only as formal
algebraic objects. One can note that in physics partition functions are
often considered which are not positive, then the sense of all calculations
is open. In some cases quantum discrete spaces can be useful, see the next
section.

Let us give some examples, see more details in \cite{ReiRov}. For a vertex $v
$ denote $H_{v}=H_{l_{1}}\otimes ...\otimes H_{l_{m}}$, where $l_{1},...l_{m}
$ are all 1-cells incident to $v$. As each 2-cell $f_{i}$, incident to $v$,
is incident to two cells emanating from $v$, then the correponding $%
H_{a(f_{i})}$ appears twice in $H_{v}$. That is why $H_{v}$\ can be
naturally represented as a tensor square, thus for each its element one can
define its trace $Tr$. The Ooguri-Turaev-Crane-Yetter model is defined by
the function 
\[
\Phi =Tr(b(l_{1})\otimes ...\otimes b(l_{m}))
\]
The corresponding partition function is finite if as a group the quantum
group $SU(2)_{q}$ is chosen. The main result is that $Z(\Gamma )$ does not
depend on $\Gamma $, if $\Gamma $ runs through 2-dimensional skeletons of
triangulations of some fixed 4-dimensional manifold. 3-dimensional analog is
the Turaev-Viro model, and 3-dimensional analog with the Lee group $SU(2)$
is the Ponzano-Regge model (see \cite{Wit}), which develops the initial
Regge model and assumes the discreteness of the lengths of edges: $%
l=\varepsilon j,j=0,\frac{1}{2},1,\frac{3}{2},...$. With each $j$ the
representation of $SU(2)$ having spin $j$ is associated.

Spin foam models give a possibility for the Gibbs families approach to the
so called loop quantum gravity, that was developed earlier in the
hamiltonian framework. It was based on the so called new variables,
introduced by Ashtekar in the classical general relativity, and used Penrose
networks as a convenient basis in the space of functions on the set of
connections on the fixed time slice $M$ of the space-time represented as $%
M\times R$.

\paragraph{Topological quantum field theory}

Axioms of topological quantum field theory as introduced in \cite{Ati}
resemble more axioms of $S$-matrix theory, than give models of the
space-time structure. Gibbs families then reflect the well-known physical
presentation of topological field theory in terms of functional integrals.
The known models of TFT deal with finite dimensional Hilbert spaces
associated with boundary manifolds. Some classification of such cases in
dimension 2 can be seen in the last chapter of \cite{amdujo}. Here we can
only relate this to some problems for Gibbs families.

Consider triangulations of a two dimensional compact oriented surface of
genus $\rho $ with $k$ holes, let $S$ be the spin space. Let $\mathcal{T}%
(N,m_{1},...,m_{k})$ be the set of triangulations with fixed number $N$ of
triangles and given numbers $m_{1},...m_{k}$ of edges on each of $k$
boundaries. For given potential one can consider the conditional partition
function $Z_{m_{1}...m_{k}}(s_{boundary})$ given the values $s_{boundary}$
of the spin on the boundary, this gives a positive measure $\mu
(m_{1},...,m_{k})$ (non-normalized) on the configurations $S^{m_{1}}\times
...\times S^{m_{k}}$ and a measure on 
\[
D(k)=\cup _{m_{1},...,m_{k}}S^{m_{1}}\times ...\times S^{m_{k}}
\]
\ for given $k$, and even on $\cup _{k}D(k)$. Most questions about these
measures are open. In particular, when these measures are finite and under
what scalings the weak limit of the normalized measures exists ? This can
give examples when the mentioned Hilbert spaces have infinite dimension.

\paragraph{Causal sets}

Causal set (see short review \cite{Rei}) $V$, that is a partially-ordered
set with relation $\leq $, satisfying transitivity, reflexivity and
acyclicity (if $x\leq y$ and $y\leq x$ then $x=y$). Given (finite or
countable) causal set defines a directed graph $G$ (Hasse diagram of $V$),
where $V$ is the set of vertices of $G$, and there is a directed edge from $x
$ to $y$, if $x\leq y,x\neq y$ and there is no other $z$ such that $x\leq
z\leq y$. Vice versa, each directed graph $G$ without cycles defines a
causal set $V=V(G)$. \ Thus causal sets can be reduces to graphs with a
local structure. Causal set models are useful because for them discrete
analogs of future and light cones and other conceipts, used in the classical
general relativity, can be defined.

\paragraph{Noncommutative lattices}

To a noncommutative $C^{\ast }$-algebra corresponds the dual object,
interpreted as a discrete noncommutative space. Strictly speaking this
topics is related to the second part of this paper, and we shall say only
some words.

Topological space $X$ can be approximated by discrete sets in two ways:
either one takes a sequence of denumerable subsets $X_{1}\subset ...\subset
X_{n}\subset ...\subset X$, more and more dense, or use a system of embedded
coverings $\mathcal{U}^{n}=\left\{ \mathcal{U}_{i}^{n}\right\} $ (and
consider the set of all closed points of its projective limit). Arbitrary
covering of a compact space $X$ generates a finite partition $X$, moreover
the set of blocks of this partition will be a $T_{0}$-space in the factor
topology. $T_{0}$-topology on a finite or countable set is equivalent to
some partial order on it: $x\leq y$ iff $y$ belongs to the closure of $%
\left\{ x\right\} $. As we have seen, with partially ordered sets a directed
graph is naturally associated - its Hasse diagram.

Each finite partially ordered set $V$ is the set of irreducible
representations $\hat{A}$ of some (nonunique in general) $C^{\ast }$-algebra 
$A$, see \cite{Landi}. If moreover $A$\ is separable, then the set $\hat{A}$%
\ is homeomorphic to the set $primA$ of primitive ideals of $A$, which is
called a noncommutative lattice. Application of noncommutative lattices to a
simple quantum mechanical problem see in the last chapter of \cite{Landi}.

\section{Quantum discrete spaces}

\subsection{Quantum graphs}

Let $S$ be finite or countable. Denote $\mathcal{H}=l_{2}(\mathcal{A}^{(0)})$
the Hilbert space with the orthonormal basis $e_{\alpha }$, enumerated by
finite spin graphs $\alpha \in \mathcal{A}^{(0)}:(e_{\alpha },e_{\beta
})=\delta _{\alpha \beta }$, where the function$e_{\alpha }(\beta )=\delta
_{\alpha \beta }$. Each vector $\phi $ from $\mathcal{H}$ is a complex
function on the set $\mathcal{A}^{(0)}$ and can be written as 
\[
\phi =\sum \phi (\alpha )e_{\alpha }\in \mathcal{H},\left\| \phi \right\|
^{2}=\sum \left| \phi (\alpha )\right| ^{2} 
\]
The states of the systems (wave function) are vectors $\phi $ with $\left\|
\phi \right\| ^{2}=\sum \left| \phi (\alpha )\right| ^{2}=1$.

The definition of quantum ''Gibbs'' states differs from classical in the
fact that in quantum cases dynamics, or the hamiltonian, emerges
automatically. Dynamics on quantum discrete space is associated with local
substitutions.\ Roughly speaking, a subsitution consists in the change of
one regular subgraph $\gamma $ with another one $\delta $, but the exact
definition is somewhat laborous because one should take care about rules to
connect $\delta $ with the complement of $\gamma $.

Vertex $v\in V(\gamma )\subset V(G)$ of a regular subgraph $\gamma $ of $G$
is called internal (to $\gamma $), if any of its incident edges connects $v$
with another vertex of $\gamma $, and call it boundary otherwise. The set of
all boundary edges call the boundary $\partial \gamma $ of the regular
subgraph $\gamma $.

\begin{definition}
The substitution rule (production) $Sub=(\Gamma ,\Gamma ^{\prime
},V_{0},V_{0}^{\prime },\varphi )$ is defined by two ''small'' spin graphs $%
\Gamma $ and $\Gamma ^{\prime }$, two subsets $V_{0}\subset V=V(\Gamma
),V_{0}^{\prime }\subset V^{\prime }=V(\Gamma ^{\prime })$ and one-to-one
mapping $\varphi :V_{0}\rightarrow V_{0}^{\prime }$, respecting the spins.
Let an isomorphism $\psi :\Gamma \rightarrow \gamma $ onto some \emph{regular%
} spin subgraph $\gamma $ of the spin graph $\alpha $ be given, such that $%
\psi (V_{0})\supset \partial \gamma $. The transformation (substitution) $%
T=T(Sub,\psi )$ on the spin graph $\alpha $, corresponding to the given
substitution rule $Sub$ and to the isomorphism $\psi $, is defined in the
following way. Delete all edges of the subgraph $\gamma =\psi (\Gamma )$,
delete all vertices of $V(\gamma )\setminus \psi (V_{0})$, in the non
connected union of $\alpha $ and $\Gamma ^{\prime }$ identify each $\psi
(v),v\in V_{0},$ with $\varphi (v)\in V_{0}^{\prime }$. Denote the resulting
spin graph $T(Sub,\psi )\alpha $. The function $s$ on $V(\alpha )\setminus
V(\gamma )$ is inherited from $\alpha $, and on $V(\gamma )$ - from $\Gamma
^{\prime }$.
\end{definition}

Examples of substitutions: deletion of an edge, appending an edge in a
vertex with a new vertex, connecting two vertices with an edge, changing the
spin in one vertex. The mere possibility of such substitution may depend on
some neighborhood of the vertex.

For any $N$ and $p$ let $\mathcal{H}_{N,p}\subset \mathcal{H}$ be a finite
dimensional space, generated by all $e_{\alpha },\alpha \in \cup _{n=0}^{N}%
\mathcal{A}_{n,p}^{(0)}$, let $P_{N}$ be the orthogonal projector on $%
\mathcal{H}_{N,p}$. For a given $p$ there are natural imbeddings $\mathcal{H}%
_{N,p}\subset \mathcal{H}_{N+1,p}$. Further on $p$ is fixed and we omit this
index, having in mind that degrees of vertices cannot exceed $p$. Denote $%
\mathcal{H}_{N}=\mathcal{H}_{N,p}$.

\begin{definition}
A grammar (more exactly grammar on graphs) is defined by a finite set of
substitution rules $Sub_{i}=(\Gamma _{i},\Gamma _{i}^{\prime
},V_{i,0},V_{i,0}^{\prime },\varphi _{i}),i=1,...,\left| Sub\right| $. It is
called local, if for all $i$ the graphs $\Gamma _{i}$, corresponding to $%
Sub_{i}$, are connected. \ We call grammar locally bounded if for $p$
sufficiently large the sets $\cup _{N}\mathcal{A}_{N,p}^{(0)}$ are invariant
with respect to any of its substitutions.
\end{definition}

Define the main operators $a_{i}(j)$. Index $i$ corresponds to one of its
substitution rules of the grammar. For a given $i$ consider all possible
pairs $(\psi ,\xi )$, where $\xi \in \mathcal{A}^{(0)}$, and $\psi $ is a
mapping of $\Gamma _{i}$ onto a regular subgraph of the spin graph $\xi $.
The pair $(\psi ,\xi )$\ is called minimal, if $\psi (\Gamma _{i})$ does not
belong to a neighborhood of $0$ of radius less than the radius $R(\xi
)=R_{0}(\xi )$ of the graph $\xi $. For given $i$ enumerate all possible
regular pairs with indices $j=1,2,...$%
\[
(\psi _{i1},\xi _{i1}),(\psi _{i2},\xi _{i2}),...,(\psi _{ij},\xi _{ij}),... 
\]
If in $\xi $ there are at least twp regular subgraphs, isomorphic to $\Gamma
_{i}$ and coming one to another under the action of a nontrivial
automorphism of the graph $\xi $, then they have different indices inj our
numeration.

Put 
\[
a_{i}(j)e_{\alpha }=e_{\beta },\beta =T(Sub_{i},\rho \psi _{ij})\alpha 
\]
for any $\alpha $ such that there exists an isomorphism $\rho $ of the spin
graph $\xi _{ij}$ onto $O_{R(\xi )}(0)$ of the origin of $\alpha $. Put $%
a_{i}(j)e_{\alpha }=0$ \ otherwise. It is clear that this definition does
not depend on the choice of $\rho $. Note that $\left\| a_{i}(j)\right\| =1$.

Define the linear operator $H=H(\left\{ Sub_{i},i=1,...,k\right\} $,
corresponding to the grammar $Sub_{i}=(\Gamma _{i},\Gamma _{i}^{\prime
},V_{i,0},V_{i,0}^{\prime },\varphi _{i}),i=1,...,k$, as the formal sum 
\[
H=\sum_{i=1}^{k}\sum_{j}\lambda _{i}a_{i}(j) 
\]
for some complex constants $\lambda _{i}$. Then $H$ is defined on the linear
space $\mathcal{H}^{0}$ of finite linear combinations of $e_{\alpha }$. It
is important to note that this operator does not depend on the enumeration $%
\psi _{i,j}$ due to the following equivalent definition: for all $\alpha $ 
\[
He_{\alpha }=\sum_{i=1}^{k}\sum_{\psi }e_{T(Sub_{i},\psi )\alpha } 
\]
where for given $i$ the sum over all imbeddings $\psi :\Gamma
_{i}\rightarrow \alpha $.

Adjoint substitution rule $Sub^{\ast }=(\Gamma ^{\ast },\Gamma ^{\prime \ast
},V_{0}^{\ast },V_{0}^{\prime \ast },\varphi ^{\ast })$ to the substitution
rule $Sub$ is defined by the properties: $\Gamma ^{\ast }=\Gamma ^{\prime
},\Gamma ^{\prime \ast }=\Gamma ,V_{0}^{\ast }=V_{0}^{\prime },V_{0}^{\prime
\ast }=V_{0},\varphi ^{\ast }=\varphi ^{-1}$. If substitutions $i$ and $%
i_{1} $ are adjoint then for any $j$ there exists such $j_{1}$ that $%
(a_{i}(j))^{\ast }=a_{i_{1}}(j_{1})$.

The hamiltonian is formally symmetric, $H=H^{\ast }$, if for each $i$ there
exists $j$ that $Sub_{j}$ is adjoint to $Sub_{i}$ and $\lambda _{i}=\bar{%
\lambda}_{j}$.

\subparagraph{Grammars for linear graphs}

If $p=2$ then the graph is linear, cyclic or not. Equivalently one can say
about a word $\alpha =x_{1}...x_{n},x_{i}\in S$. In this case any
substitution rule is a pair $\gamma \rightarrow \delta $, and substitutions
are transformation of words as 
\[
\alpha \gamma \beta \rightarrow \alpha \delta \beta 
\]
for any $\alpha ,\beta $, where the concatenation of $\alpha =x_{1}...x_{n}$
and $\beta =y_{1}...y_{m}$ is defined as 
\[
\alpha \beta =x_{1}...x_{n}y_{1}...y_{m} 
\]
The operator $a_{i}(j)$ allows to do the substitution $i$ in the word $%
\alpha $ only for a factor (subword) $\gamma _{i}$, beginning with the
symbol with number $j$. Adjoint to the substitution rule $\gamma \rightarrow
\delta $ is the substitution rule $\delta \rightarrow \gamma $.

\subparagraph{Selfadjointness}

Let $H$ be symmetric on the linear space $\mathcal{H}^{0}$ of finite linear
combinations of $e_{\alpha }$. They are $C^{\infty }$-vectors (see \cite
{resi}) for $H$, that is $He_{\alpha }\in \mathcal{H}^{0}$. Note that $H$ is
unbounded in general.

\begin{theorem}
Let the grammar be locally bounded and $H$\ be symmetric. Then $H$ is
essentially sekfadjoint on $\mathcal{H}^{0}$.
\end{theorem}

One can get the proof from the following lemma and Nelson's criteria, see 
\cite{resi}. This defines the unitary group $\exp (itH)$. Note that in the
lemma one does not assume symmetricity of the hamiltonian.

\begin{lemma}
For any grammar, a vector $\phi \in \mathcal{H}^{0}$ is analytic for the
corresponding hamiltonian, that is 
\[
\sum_{k=0}^{\infty }\frac{\left\| H^{k}\phi \right\| }{k!}t^{k}<\infty 
\]
for some $t>0$.
\end{lemma}

Proof. It is sufficient to take $\phi =e_{\alpha }$ for some $\alpha $. In
this case the number of pairs $(i,j)$ such that $a_{i}(j)e_{\alpha }\neq 0$,
does not exceed $CV,V=V(\alpha )$, for some constant $C$. In fact, the
number of regular subgraphs, isomorphic to a given ''small'' subgraph,
depends on the number of vertices and on $p$. Write the expansion of $H$ as 
\[
H=\sum_{a}V_{a} 
\]
where $V_{a}$ is equal to some of $\lambda _{i}a_{i}(j)$. Then 
\begin{equation}
H^{n}e_{\alpha }=\sum_{a_{n},...,a_{1}}V_{a_{n}}...V_{a_{1}}e_{\alpha }=\sum
C_{\beta }e_{\beta }  \label{exp11}
\end{equation}
Maximal number of vertices of graphs $\beta $ in the expansion $%
V_{a_{n}}...V_{a_{1}}e_{\alpha }$ does not exceed $V(\alpha )+C_{1}n$, as
any factor $V_{a_{j}}$ increases the number of vertices on not more than
some constant $C_{1}$. Then for given $e_{\alpha },a_{1},...,a_{n}$ the
nyumber of operators $V_{a_{n+1}}$, giving nonzero contribution to $%
V_{a_{n+1}}V_{a_{n}}...V_{a_{1}}e_{\alpha }$ is not greater than $C(V(\alpha
)+C_{1}n)$ for some constant $C$. Thus, the number of nonzero terms $%
V_{a_{n}}...V_{a_{1}}e_{\alpha }$ is not greater than 
\[
C^{n}\prod_{j=1}^{n}(\left| V(\alpha )\right|
+C_{1}j)=(CC_{1})^{n}\prod_{j=1}^{n}(\frac{\left| V(\alpha )\right| }{C_{1}}%
+j)< 
\]
\[
<(CC_{1})^{n}n!\prod_{j=1}^{n}(1+\frac{\left| V(\alpha )\right| }{C_{1}j}%
)<(CC_{1})^{n}n!n^{2\frac{\left| V(\alpha )\right| }{C_{1}}} 
\]
and the norm of each term is bounded by $\left( \max \lambda _{i}\right)
^{n} $. This gives convergence of the series for $\left| t\right| <t_{0}$,
where $t_{0}$ does not depend on $\alpha $. Lemma is proved.

\subsubsection{$C^{\ast }$-algebras}

There are several useful $C^{\ast }$-algebras, related to grammars on
graphs: two universal $\mathbf{B},\mathbf{C}$ and one $\mathbf{A}(Gr)$\
depending on the grammar. Define these algebras.

Let $\mathbf{C}_{N}$\ be the $C^{\ast }$-algebra of all operators in the
finite dimensional space $\mathcal{H}_{N}=\mathcal{H}_{N,p}$. The operator $%
C\in \mathbf{C}_{N}$ can be considered as acting in $\mathcal{H}$, if we put 
$Ce_{\alpha }=0$ for $e_{\alpha }\bar{\in}\mathcal{H}_{N,p}$. Then $\mathbf{C%
}_{N}\subset \mathbf{C}_{N+1}$, put\textbf{\ }$\mathbf{C}=\overline{\cup _{N}%
\mathbf{C}_{N}}$. The algebra $\mathbf{C}$ consists of compact operators. It
was used in in \cite{m4}.

Let $\mathbf{A}_{N}=\mathbf{A}_{N}(Sub_{i},i=1,...,k)$ be the $C^{\ast }$%
-algebra of operators in $\mathcal{H}$, generated by all $a_{i}(j)$ such
that $R(\xi _{ij})\leq N$. Then the natural imbeddings $\phi _{N}:\mathbf{A}%
_{N}\rightarrow \mathbf{A}_{N+1}$ are defined. The inductive limit $\cup _{N}%
\mathbf{A}_{N}=\mathbf{A}^{0}$ of $C^{\ast }$-algebras $\mathbf{A}_{N}$ is
called the local algebra and its norm closure $\mathbf{A=A}%
(Sub_{i},i=1,...,k)$ is called the quasilocal algebra. Structure of the \ $%
C^{\ast }$-algebra $\mathbf{A=A}(Sub_{i},i=1,...,k)$\ depends on the grammar
and on the fixed class of graph. Below we give many bexamples of
hamiltonians and $C^{\ast }$-algebras.

Universal $C^{\ast }$-algebra $\mathbf{B}$ is generated by the operators $%
a_{i}(j)$ for all possible substitutions. $\mathbf{B}$ does not coincide
with the algebra of all bounded operators, for example, many diagonal
operators in the basis $e_{\alpha }$ do not belong to $\mathbf{B}$.

\subsubsection{Automorphisms}

Further on the local boundedness of the grammar is always assumed and we fix 
$p$ such that the sets $\cup _{N}\mathcal{A}_{N,p}^{(0)}$ are invariant with
respect to all substitutions of the grammar.

Formal hamiltonian defines a differentiation in the local algebra $\mathbf{A}%
^{0}$. Consider 
\[
H_{N}=\sum_{i=1}^{\left| Sub\right| }\sum_{j:R(\xi _{ij})\leq N}\lambda
_{i}a_{i}(j) 
\]
Define the automorphism group $\mathbf{A}$ as follows. Take a local element $%
A\in \mathbf{A}^{0}$ and $N$ such that $A\in \mathbf{A}_{N}$, and put 
\[
\alpha _{t}^{(N)}(A)=\exp (iH_{N}t)A\exp (-iH_{N}t) 
\]

\begin{theorem}
There exists $t_{0}>0$ such that for any local $A$ and any $t,\left|
t\right| <t_{0}$, there exists the norm limit 
\[
\lim_{N\rightarrow \infty }\alpha _{t}^{(N)}(A) 
\]
This defines the unique automorphism of the quasilocal algebra.
\end{theorem}

Proof. Consider the Dyson-Schwinger series 
\[
A_{t}^{(N)}=A+\sum_{n=1}^{\infty }\frac{(it)^{n}}{n!}[%
H_{N},...,[H_{N},[H_{N},A]]...] 
\]
One can restrict himself to local $A=a_{i}(k)$ for some $i,k$.

\begin{lemma}
The norm of the operator $\frac{(it)^{n}}{n!}[H_{N},...,[H_{N},[H_{N},A]]...%
] $ is bounded from below by $(Ct)^{n}$ independently of $N$.
\end{lemma}

Note that the commutator belongs to $\mathbf{A}$\ and is equal to the sum of
commutators 
\[
\lbrack a_{i_{n}}(j_{n}),...,[a_{i_{2}}(j_{2}),[a_{i_{1}}(j_{1}),A]],...] 
\]
multiplied with $\lambda _{i_{1}}...\lambda _{i_{n}}$. Each commutator is
equal to the sum of $2^{n}$ terms like 
\[
\pm a_{q_{n}}(p_{n})...a_{i_{q2}}(p_{2})a_{q_{1}}(p_{1}) 
\]
Let us prove that after cancellations there will be not more than $%
\prod_{j=1}^{n}Cj=C^{n}n!$ terms left (independently of $N$). One should use
for this the second definition of the hamiltonian. Introduce the following
notation: if $a_{i}(j)e_{\alpha }=e_{T(Sub,\psi )}$, then we shall write $%
a_{i}(j)e_{\alpha }=a(T(Sub,\psi ))e_{\alpha }$. Then for any $\alpha $, any
two substitution rules $i_{1},i_{2}$ and two mappings $\psi _{1}:\Gamma
_{i_{1}}\rightarrow \gamma _{1},\psi _{2}:\Gamma _{i_{2}}\rightarrow \gamma
_{2}$ onto non-intersecting regular subgraphs $\gamma _{1},\gamma _{2}$ of
the graph $\alpha $, the corresponding transformations in terms of $a_{i}(j)$
will give cancellating terms as 
\[
a(T(T(Sub_{1},\psi _{1})\alpha ,Sub_{2},\psi _{2}))a(T(\alpha ,Sub_{1},\psi
_{1}))e_{\alpha }-a(T(T(Sub_{2},\psi _{2})\alpha ,Sub_{1},\psi
_{1}))a(T(\alpha ,Sub_{2},\psi _{2}))e_{\alpha }=0 
\]
etc. by induction.

Similarly one can prove that for $N\rightarrow \infty $ $A_{t}^{(N)}$
converges term-by-term (for any $n$) to the series 
\[
A_{t}=A+\sum_{n=1}^{\infty }\frac{(it)^{n}}{n!}[H,...,[H,[H,A]]...] 
\]
Each term of the this series is defined and the series converges, by this
estimate, in norm for sufficiently small $t$. The existence of automorphism
group fro all $t$ can be proved similrly to the Robinson theorem for quantum
spin systems, see \cite{brro}.

\subsubsection{KMS-states on $\mathbf{A}$}

We shall consider KMS states on $\mathbf{A}$, which are a natural
generalization of KMS states on the lattice. Sometimes it is useful however
to consider KMS states on the algebra $\mathbf{C}$ (see \cite{m4}), where
the situation is a bit different.

We restrict ourselves with the case $p=2$, that is with linear graphs and
assume that $2\leq \left| S\right| <\infty $. The algebra $\mathbf{A}_{N}$,
contrary to quantum spin systems, is not finite dimensional in general (if
the length of the word can change by substitution). One can introduce finite
dimensional algebras $\mathbf{F}_{N}$, generated by all operators $%
a_{i}^{(N)}(j)=P_{N}a_{i}(j)P_{N}$. One can assume that all these operators
act in $\mathcal{H}_{N}$. Define states on $\mathbf{F}_{N}$\ by putting for
any $A_{N}\in \mathbf{F}_{N}$ 
\[
<A_{N}>_{N}=Z_{N}^{-1}Tr_{N}\left[ A_{N}\exp (-\beta P_{N}HP_{N})\right] 
\]
\[
Z_{N}=Tr_{N}\exp (-\beta P_{N}HP_{N}) 
\]
where $Tr_{N}$ is the trace in $\mathcal{H}_{N}$. Any limiting point of this
sequence defines a state on $\mathbf{A}$%
\[
<A>=\lim_{N\rightarrow \infty }<P_{N}AP_{N}>_{N} 
\]
where $A$ is local.

\begin{theorem}
If $\beta $ is sufficiently small then $\log Z_{N}\sim fN$ for some constant 
$f>0$.
\end{theorem}

Proof. Note first that for $H=0$ this is evident as $Tr_{H}1=\dim \mathcal{H}%
_{N}=2^{N+1}$.

One can write 
\[
Z_{N}=\sum_{\alpha :R_{0}(\alpha )\leq N}z(\alpha ),z(\alpha )=(e_{\alpha
},\exp (-\beta P_{N}HP_{N})e_{\alpha }) 
\]
or 
\begin{equation}
Z_{N}=\sum_{k=0}^{\infty }\sum_{I_{k},J_{k}}\sum_{\alpha :\left| \alpha
\right| \leq N}\frac{(-\beta )^{k}}{k!}z(\alpha ,(I_{k},J_{k}))  \label{exp1}
\end{equation}
where for ordered arrays $I_{k}=(i_{1},...,i_{k}),J_{k}=(j_{1},...,j_{k})$ 
\[
z(\alpha ,(I_{k},J_{k}))=\lambda _{i_{1}}...\lambda _{i_{k}}(e_{\alpha
},P_{N}a_{i_{k}}(j_{k})P_{N}...P_{N}a_{i_{1}}(j_{1})P_{N}e_{\alpha }), 
\]
that is $z(\alpha ,(I_{k},J_{k}))$ equals either $\lambda _{i_{1}}...\lambda
_{i_{k}}$\ or $0$. Our goal is to obtain a cluster expansion for 
\[
z(N)\doteq \sum_{\alpha :\left| \alpha \right| =N}z(\alpha ) 
\]
of the kind 
\begin{equation}
z(N)=\sum c_{U_{1}}...c_{U_{n}}  \label{exp2}
\end{equation}
which is defined exactly below, to make it possible to apply the general
theory of cluster expansions from \cite{malmin1}. To get the xpansion \ref
{exp2} we use resummation of the expansion \ref{exp1}. To get \ref{exp2} we
need an inductive construction.

In \ref{exp2} the sum is taken over arrays of non-intersecting subsets $%
(U_{1},...,U_{n})$. Denote $U_{0}=\left\{ 1,...,N\right\} \setminus \cup
_{i=1}^{n}U_{i}$. We define first $U_{0}$, and thendefine the ''clusters'' $%
U_{i}$ and numbers $c_{U_{i}}$.

We shall do it separately for each term 
\[
\frac{(-\beta )^{k}}{k!}z(\alpha ,(I_{k},J_{k})) 
\]
corresponding to the word $\alpha $ of length $N$ and some $(I_{k},J_{k})$,
after this we resumm the results.

\subparagraph{Definition of $U_{0}=U_{0}(\protect\alpha ,(I_{k},J_{k}))$.}

We say that the substitution $T=T(Sub,\psi )$ does not touch symbol on the
place $v$ of the word $\alpha $, if the image $\psi $ does not contain $v$.
It is convenient to denote $T_{p}$ the substitution corresponding to the
operator $b(p)=a_{i_{p}}(j_{p}),p=1,...,k$. The symbol $v$ of the word $%
\alpha $ is untouched for given $(I_{k},J_{k})$, if none of the
substitutions $T_{p},p=1,...,k,$ does not touch it. Similarly, the symbol $v$
of the word $b(s)...b(1)e_{\alpha }$ is called untouched for given $%
(I_{k},J_{k})$ if none of $T_{l}$ with $s<l\leq k$ does not touch it. For
given $(\alpha ,(I_{k},J_{k}))$ the set $U_{0}=U_{0}(\alpha ,(I_{k},J_{k}))$
is defined as the set of all symbols of the word $\alpha $ belonging to all
words $b(s)...b(1)e_{\alpha }$ and not touched for all $s$.

\subparagraph{Definition of clusters $U_{i}(\protect\alpha ,(I_{k},J_{k}))$.}

Denote $\alpha _{0}=\alpha ,\alpha _{s}=T_{s}...T_{1}\alpha ,s\geq 1$.
Introduce the system of partitions $g_{sr},r,s=0,...,k$. of the set of
symbols of the word $\alpha _{s}$ into subset. The partition $g_{0}=g_{0,0}$
is defined as the partition of $\alpha $ on separate symbols, that is the
partition of $\left\{ 1,...,N\right\} $ on $N$ blocks.

For a given substitution $T$, that trasforms the word $\alpha $ to the word $%
\beta $, and for a given partition $g=g(\alpha )$ of the set $V(\alpha )$ of
symbols of the word $\alpha $ define partitions $g(\beta )$ of the word $%
\beta $.\ Let $V_{1}(\alpha ,T)$ be the set of symbols of the word $\alpha $%
, left untouched by the substitution $T$ . Assume that the substitution $T$
deletes in the new word $\beta $\ the set $V(\alpha )\setminus V_{1}(\alpha
) $ of symbols, and appends some set $V_{2}$ of symbols. Thus, $V(\beta
)=V_{1}(\alpha )\cup V_{2}$. Let us call the partition $g(\beta )=g(\beta
,g(\alpha ),T)$ of the word $\beta $ the partition induced by the partition $%
g=g(\alpha )$ and by the substitution $T$ \ if the following conditions
hold: 1) Is the block $I$ of the partition $g(\alpha )$ belongs to $%
V_{1}(\alpha )$, then this block will be again a block of the partition $%
g(\beta )$; 2) The vertices of $V_{2}$ build one block together with all
vertices of other blocks $I$ (not belonging to $V_{1}(\alpha )$) of the
partition $g(\alpha )$, having nonempty intersection with $V(\alpha
)\setminus V_{1}(\alpha )$.

Define now by induction the partitions $g_{s}=g_{s,0}=g_{s}(\alpha _{s})$ of
the words $\alpha _{s}$. If $g_{s}$ is already defined \ then $g_{s+1}$ is
the partition of the word $\alpha _{s+1}$ induced by the substitution $%
T_{s+1}$ and by the partition $g_{s}$ of the word $\alpha _{s}$. Further on
we use induction in $r$. If the partition $g_{s+1,r}$ of the word $\alpha
_{s+1}$ is defined then $g_{s,r+1}$ is defined as the partition of the word $%
\alpha _{s}$ induced by the adjoint substitution $T_{s}^{\ast }$.

Consider the partition $g_{0,k}$. Its blocks are some untouched separate
symbols as well as others, consisting of more than one symbol. These others
we denote by $U_{i}=U_{i}(\alpha ,(I_{k},J_{k}))$ and call clusters with
respect to a given pair $\alpha ,(I_{k},J_{k})$. Note that the number $%
L=L(\alpha ,(I_{k},J_{k}))$ of clusters cannot exceed $k$. Moreover, the set 
$\left\{ 1,...,k\right\} $ is uniquely partitioned on the sets $%
M_{1}=M_{1}(\alpha ,(I_{k},J_{k})),...,M_{L}=M_{L}(\alpha ,(I_{k},J_{k}))$
of permutations related to the clusters $1,...,L$ correspondinglyt. Let $%
(I_{m_{i}},J_{m_{i}}),I_{m_{i}}=$ $%
I_{m_{i}}(I_{k},J_{k}),J_{m_{i}}=J_{m_{i}}(I_{k},J_{k})$ be the subarray
(with the induced order) of the array $(I_{k},J_{k})$, corresponding to $%
M_{i}$.

\subparagraph{Definition of the expansion}

Consider some set $B\subset \left\{ 1,...,N\right\} $. If some cluster can
correspond to it then it is an interval.

Let $M_{1},...,M_{L}$ be subsets of the set $\left\{ 1,...,k\right\} $ such
that $\cup _{i=1}^{L}M_{i}=\left\{ 1,...,k\right\} $. Denote $m_{i}$ the
number of elements in $M_{i}$. To get a multiplicative expansion \ref{exp2}
take the sum of all $\frac{(-\beta )^{k}}{k!}z(\alpha ,(I_{k}^{\prime
},J_{k}^{\prime }))$, where the arrays $(I_{k}^{\prime },J_{k}^{\prime })$
differ from $(I_{k},J_{k})$ only in the order of terms, moreover the order
inside each of the sets $M_{i}$ does not change. This sum is equal to 
\[
\frac{(-\beta )^{k}}{m_{1}!...m_{L}!}z(\alpha ,(I_{k},J_{k}))=\prod_{i=1}^{L}%
\frac{(-\beta )^{m_{i}}}{m_{i}!}z(\alpha
,(I_{m_{i}}((I_{k},J_{k})),J_{m_{i}}((I_{k},J_{k}))))= 
\]
\[
==\prod_{i=1}^{L}\frac{(-\beta )^{m_{i}}}{m_{i}!}\prod_{j\in M_{i}}\lambda
_{j} 
\]
To define the expansion \ref{exp2} we use resummation. For given subsets $%
U_{0},U_{1},...,U_{n}$ consider all $z(\alpha ,(I_{k},J_{k}))$ such that $%
U_{i}(z(\alpha ,(I_{k},J_{k})))=U_{i},i=0,1,...,n$, and define for a given
connected $U$%
\[
c_{U}=\sum \frac{(-\beta )^{k}}{k!}z(\alpha ,(I_{k},J_{k})) 
\]
where the sum is over all words $\alpha $ with the set $V(\alpha )=U$ of
symbols, and over all arrays $(I_{k},J_{k})$, for which $U$ is one connected
cluster. We have the following ''cluster estimate'' 
\[
c_{U}<c_{1}(C\beta )^{|U|} 
\]
By general cluster expansion theory this gives $\log z(N)\sim fN$ for some
constant $f>0$. Thus 
\[
\log Z_{N}\sim \log \left[ z(N)(1+\frac{z(N-1)}{z(N)}+...\right] \sim cN 
\]

\begin{theorem}
There exists $\beta _{0}>0$ such that for any local $A$ the limit 
\[
<A>=\lim_{N\rightarrow \infty }<P_{N}AP_{N}>_{N} 
\]
exists and is analytic in $\beta $ for $\beta <\beta _{0}$.
\end{theorem}

Proof. Take for example $A=P_{j,\delta }$, that is the projector on the
subspace generated by words containing thye word $\delta $ on place $j$. We
have 
\[
Tr_{N}\left[ \exp (-\beta P_{N}HP_{N})A\right] =\sum_{k=0}^{\infty }\frac{%
(-\beta )^{k}}{k!}Tr_{N}((P_{N}HP_{N})^{k}A)=\sum_{k=0}^{\infty }\frac{%
(-\beta )^{k}}{k!}%
\sum_{I_{k},J_{k}}Tr_{N}(a_{i_{k}}(j_{k})...a_{i_{1}}(j_{1})A) 
\]
where the sum is over all $I_{k}=(i_{1},...,i_{k}),J_{k}=(j_{1},...,j_{k})$.
The rest of the proof is quite standard and uses the genral cluster
expansion theory, we omit it.

\begin{remark}
The case of spin graphs with $p\geq 3$ should be similar to the theory
developped in the section 4.3.
\end{remark}

\subsection{Examples and structure of the hamiltonians}

From general theory now we come to examples for linear graphs. But first we
want to show connections with Hooft quantization.

\subsubsection{Hooft quantization}

Already Feynman \cite{feyn}\ suggested that on the Planck scale physical
laws can resemble transformations of finite automata with discrete time. In
the series of papers \cite{hooft1, hooft2} Hooft discusses quantization of
discrete (deterministic) mappings.

If $f:S\rightarrow S$ is a mapping of the finite set $S$ to itself, then
there exists a subset $S_{0}$, invariant with respect to $f$, moreover $%
f:S_{0}\rightarrow S_{0}$ is one-to-one. In $l_{2}(S_{0})$ to this mappin,g
corresponds some permutation matrix $U$, defined by 
\[
(U\phi )(s)=\phi (f^{-1}s) 
\]
The transformation $U$ in the Hilbert space $l_{2}(S)$ is defined as the
quantization (in the sense of Hooft) of the deterministic transformation $f$%
. Moreover, $U$ is unitary equivalent to the sum of cyclic shifts on
discrete circles. Examples of physical systems, for which such
representation is possible, are based in fact on the unitary equivalence of
the hamiltonian to the multiplication operator, or by Fourier transform to
the shift operator.

One should say that this is a quantization of ''first order systems''. Then
there exists a matrix $H$ such that 
\[
U_{t}=\exp (itH),U_{1}=U 
\]
In noninteger time moments $U_{t}$ does not define a deterministic
transformation of $S$ to itself. Moreover, $H$ is not uniqely defined but
only up the multiplication of each eigenvalue on $\sqrt[q]{1}$ where $q$ is
length of the corresponding cycle.

The simplest physical example is: a spinning particle in the magnetic field:
in the finite dimensional space with the basis $e_{m},m=-l,...,l$, the
diagonalization of the hamiltonian is $He_{m}=\mu me_{m}$. However in
another basis $f_{n}=\frac{1}{\sqrt{N}}\sum_{m}\exp (-\frac{2\pi imn}{N}%
)e_{m},n=0,1,...,N-1;N=2l+1$, in discrete moments of time $t-t^{\prime }=%
\frac{2\pi k}{\mu N}$ we have the deterministic dynamics $\exp (i\frac{2\pi k%
}{\mu N}H)$ $f_{n}\rightarrow f_{n+k},\func{mod}N$. Hooft gives also more
sophisticated examples. However, one can use quantum grammars to construct
other examples of this kind.

Quantum grammars on graphs were introduced independently by the author, but
can be considered in the framework of more general ideas: the substitution
operation is not one-valued in general (it is nondeterministic, using the
language of the informatics), the set $S$ is countable, unbounded growth is
possible. Though there are many possibilities of how to work with
nondeterministic automata, the homogeneity (independence on the place in the
graph) makes the choice of the quantization unique up to a technical
realization: sequential transformation, sinchronous parallelism,
asinchronous parallelism. we have chisen the latter possibility. If there
are several possibilities, then there is a new possibilities to prescribe
weights to them.

\subsubsection{Quantum spin systems on a fixed graph}

If the hamiltonian does not change the graph but only the configurations
then we get a quantum spin system on a fixed graph. Vice versa, any quantum
spin system can be obtained as a quantum grammar on graphs, this can be done
by the necessary choice of the constants $\lambda _{i}$. In this case the
imbeddings $\phi _{N}:\mathbf{A}_{N}\rightarrow \mathbf{A}_{N+1}$ are given
by $A\rightarrow A\otimes 1\otimes ...\otimes 1$. Such algebras were studied
intensively, see \cite{brro}.

\subsubsection{Linear grammars and Toeplitz operators}

In some cases one gets interesting examples of\ $C^{\ast }$-algebras, which
appeared earlier from different problems. For example, the $C^{\ast }$%
-algebra $\mathbf{A}_{N}$, generated by the substitution $a\rightarrow aa$,
is isomorphic to the $C^{\ast }$-algebra of Toeplitz operators in $%
l_{2}(Z_{+})$, generated by shifts. To study the structure of the $C^{\ast }$%
-algebra is often a simpler problem than to study the spectrum of the
hamiltonian, but already this gives some information about the spectral
properties.

Consider the case when $S=\{a,w\}$ consists of two symbols, and the
substitutions are $aw\rightarrow w,w\rightarrow aw$. Put $\lambda
(aw\rightarrow w)=\lambda (w\rightarrow aw)=\lambda $. Such grammar is
called a right linear grammar in the informatics. The the subspaces $%
\mathcal{H}(L_{k})$, where $L_{k}$ is the set of words with exactly $k$
symbols $w$, are invariant. For example, $L_{1}=\{a^{m}w=aa...aw,m=0,1,2,...%
\}$. Call $\mathcal{H}(L_{k})$ the $k$-particle space. Let $H_{k}$ be the
restriction of $H$ on $\mathcal{H}(L_{k})$.

\begin{theorem}
(One particle spectrum) $H_{1}$ on $\mathcal{H}(L_{1})$ is unitary
equiavalent to the operator of multiplication on $\lambda (z+z^{-1})$ in $%
L_{2}(S^{1},d\nu )$, where $d\nu $ is the Lebesgue measure on the unit
circle $S^{1}$ in the complex plane $\mathbf{C}$.
\end{theorem}

$\mathcal{H}(L_{1})$ is isomorphic to $l_{2}(Z_{+})$ via $aa...aw\rightarrow
m$. With this nisomorphism $H$ becomes the Toeplitz operator in $%
l_{2}(Z_{+}) $ equal to 
\[
\lambda b+\lambda b^{\ast } 
\]
where $b$ is the left shift in $l_{2}(Z_{+})$%
\[
(bf)(m)=f(m-1),m=1,2,...,(bf)(0)=0 
\]
And the $C^{\ast }$-algebra $\mathbf{A}$ becomes the $C^{\ast }$-algebra $%
\mathbf{W}_{1}$ of Toeplitz operators, that is the $C^{\ast }$-algebra of
operators in $l_{2}(Z_{+})$ generated by $b$. Note that the commutator $%
\left[ b,b^{\ast }\right] $ is a finite-dimansional operator. More
generally, we have the following exact sequence of algebras 
\[
0\rightarrow K\rightarrow ^{j}W_{1}\rightarrow ^{\pi }W\rightarrow 0 
\]
where $K$ is the closed two--sided ideal in $W_{1}$, consisting of all
compact operators in $l_{2}(Z_{+})$, $j$ is the imbedding, $\pi $ is the
natural projection $W_{1}$ on $W_{1}/K\sim W$. In fact, the commutator of
any two elements from $W_{1}$ is compact. The last assertion can be easily
verified for monomials $b_{1}b_{2}...b_{k}$, where $b_{i}=b$ or $b^{\ast }$.
Passing to the limit this can be proved for any elements. It follows that $%
W_{1}/K$ is commutative. Put 
\[
y=\pi b,y^{\ast }=y^{-1}=\pi b^{\ast } 
\]
One can prove that the elements 
\[
\sum_{i=-k}^{k}c_{i}y^{i} 
\]
are all different and their norm is equal to $\sum \left| c_{i}\right| $, it
is sufficient to consider their action on the functions $f(m)$ equal $1$ for
some sufficiently large $m$ and zero otherwise. It follows that $W_{1}/K\sim
W$. We get from this that the continuous part of the spectrum coincides with
the spectrum of the multiplication operator.

To prove the absence of the discrete spectrum, rewrite the equation $%
Af-\lambda f=\phi $ \ as the well-known difference equations 
\[
\phi _{n}=-\lambda f_{n}+f_{n-1}+f_{n+1},n\geq 1 
\]
\[
\phi _{0}=-\lambda f_{0}+f_{1} 
\]
If $\phi _{n}\equiv 0$, then it does not have solutions from $l_{2}$.

Consider now the operator $H$ on all Hilbert space, but generated by words
with the symbol $w$ at the end. Each word with the symbol $w$ at the end can
be written as 
\[
\alpha =a^{m_{1}}wa^{m_{2}}w...a^{m_{k}}w,m_{1},m_{2},...,m_{k}=0,1,2,... 
\]
Correspondingly, each basis vector in $\mathcal{H}(L_{k})$ can be written as
the tensor product 
\[
e_{\alpha }=e_{\alpha (1)}\otimes e_{\alpha (1)}\otimes ...\otimes e_{\alpha
(1)},\alpha (i)=a^{m_{i}}w 
\]
In other words, $\mathcal{H}(L_{k})$ is isomorphic to the $k$-th tenor power
of $\mathcal{H}(L_{1})$, and the evolution is 
\[
\exp (itH_{1})\otimes ...\otimes \exp (itH_{1}) 
\]
or 
\[
H_{k}=H_{1}\otimes 1\otimes ...\otimes 1+...+1\otimes 1\otimes ...\otimes
H_{1} 
\]
We have proved the following result

\begin{theorem}
(multiparticle spectrum) The Hilbert space has the following decomposition
on invariant subspaces 
\[
\mathcal{H}(\Sigma ^{\ast })=\oplus _{k=0}^{\infty }\mathcal{H}(L_{k}),%
\mathcal{H}(L_{k})=\mathcal{H}(L_{1})\otimes ...\otimes \mathcal{H}(L_{1}) 
\]
The Hamiltonian $H_{k}$ on $\mathcal{H}(L_{k})$ is unitary equivalent to the
multilication on the function 
\[
\sum_{j=1}^{k}\lambda (z_{j}+z_{j}^{-1}) 
\]
in the space $L_{2}(S^{k},d\nu )$ of functions $f(z_{1},...,z_{k})$.
\end{theorem}

\subsubsection{Quantum space expansion and contraction}

Let $S$ consist of one symbol $a$. Consider two substituions $a\rightarrow
aa,aa\rightarrow a,\lambda _{1}=\lambda _{2}=\lambda $, and the hamiltonian 
\[
H=\lambda \sum_{j=1}^{\infty }(a_{1}(j)+a_{2}(j)) 
\]
with real $\lambda $. In this case the Hilbert space $\mathcal{H}$ is
isomorphic to $l_{2}(Z_{+})$, as $aa...a$ can be identified with its own
length minus $1$. The hamiltonian is unitary equivalent to the Jacobi matrix 
\[
(Hf)(n)=\lambda (n-1)f(n-1)+\lambda nf(n+1) 
\]
Let us call this operator a one particle operator, having in mind that the
''particle'' is associated with the quantum of one-dimensional space. The
evolution \ consists in the (quantum) expansion and contraction at each
point indepependently.

One can find generalized eigenfunctions from the the equation 
\[
(H_{1}-\lambda )f=\phi 
\]
or 
\[
\phi _{n}=nf_{n+1}+(n-1)f_{n-1}-\lambda f_{n} 
\]
Introducing the generating functions 
\[
F(z)=\sum_{n=1}^{\infty }f_{n}z^{n},\Phi (z)=\sum_{n=1}^{\infty }\phi
_{n}z^{n} 
\]
we shall get the following equation for $F(z)$ 
\[
F^{^{\prime }}-\frac{\lambda +\frac{1}{z}}{1+z^{2}}F=\frac{\Phi }{1+z^{2}} 
\]
The homogeneous equation has the following solutions 
\[
\exp (\int \frac{\lambda +\frac{1}{z}}{1+z^{2}}dz)=z\left( \frac{z+i}{z-i}%
\right) ^{\frac{1}{2}+\frac{i\lambda }{2}} 
\]
It follows for example the absence of discrete soectrum. See more details
about this model in \cite{mayaza}.

\subsubsection{Lorentzian Models and Grammars}

The Lorentzian model, see \cite{amjulo, amblol1}, is defined by the
following construction of the graph. On the cylinder $R\times S^{1}=\left\{
(x,t)\right\} $ on each line (slice) $t=0,...,N$ there are $l_{t}$ points $%
x_{1}(t),...,x_{l(t)}(t)$, in the clockwise order. Each pair of neighbors is
connected by an edge on the line $t=const$. Each point $x_{i}(t)$ is
connected by some number of edges with $k(i,t)=1,2,...,$ consecutive points
of the slice $t+1$, let $K(i,t+1)$ be the set of these points. It is assumed
that $K(i,t+1)\cap K(i+1,t+1)$\ consists of exactly one point,
correspondingly the right one and the left one of these sets. The
enumeration of the points is not important, the graph is considered up
isomorphisms, respecting the slices..To each graph $G$ corresponds the
amplitude 
\[
z(G)=\exp (\lambda \sum_{i,t}k(i,t)) 
\]
Note that $\sum_{i}k(i,t)$\ equals the number of triangles between the
slices $t$ and $t+1$. At the same time $\sum_{i,t}k(i,t)=2\sum_{t}l(t)$.

Here $k_{i}$ are any positive numbers, and the unitarity of this
transformation is not clear. However if $k_{i}\leq const$, one can construct
unitary transformation using quantum grammars.

Unitary model coincides with the quantum grammar describing expansion and
contraction of one-dimensional space. In fact, consider at time $0$ the
triangulation of the disk with $m$ edges on the boundary. The sequence of
edges on the boundary can be identified with the word $aa...a$ of length $m$%
. The action of the operator corresponding to the substitution $a\rightarrow
aa$ consists in the glueing a new triangle (having two edges outside the
disk) to the corresponding edge of the boundary, and the action of
contraction operator corresponding to $aa\rightarrow a$ consists in glueing
a new triangle to two consecutive edges of the triangle and having one edge
outside the disk.

Random analog of this quantum grammar was studied in \cite{m3}. It was shown
that this model belongs to a different universal class. It was shown that
only non-markovian dynamics can give the same universal class as for the
dynamical triangulation models.

Gibbs families approach to the Lorentzian models was studied inn detail in 
\cite{mayaza}. Important feature of Lorentzian models is their relations
with causal sets, see reviews \cite{reid, amjulo}.

\subsubsection{Particle on quantum space}

Consider the grammar with $S=\left\{ a,w\right\} $ and with the
substitutions 
\[
a\rightarrow aa,aa\rightarrow a,aw\rightarrow wa,wa\rightarrow aw 
\]
Subspace $\mathcal{H}_{1}$, generated by the words $%
a^{k}wa^{l},k,l=0,1,2,... $, with one only symbol $w$, is invariant. The
hamiltonian is a mixture of $H $ , defined above, and the discrete laplacian 
$H_{1}$ for free Schroedinger particle in $l_{2}$ on a finite subset.

\subsubsection{Two types of space quanta}

Let $S=\left\{ a,b\right\} $, consider 4 substitutions 
\[
1:a\rightarrow aa,2:aa\rightarrow a,3:b\rightarrow bb,4:bb\rightarrow b 
\]
We have here two invariant one-particle subspaces $\mathcal{H}_{a},\mathcal{H%
}_{b}$. For example, $\mathcal{H}_{a}$ is generated by the words $%
a,a^{2}=aa,...,a^{n},...$. There are two invariant two particle subspaces $%
\mathcal{H}_{ab}=\mathcal{H}_{a}\otimes \mathcal{H}_{b},\mathcal{H}_{ba}=%
\mathcal{H}_{b}\otimes \mathcal{H}_{a}$. For example, $\mathcal{H}_{ab}$ is
generated by the words $a^{k}b^{l},k,l>0$. In general, for any $n$ there are
two inavriant $2n$-particle subspaces: $\mathcal{H}_{(ab)_{n}}$, generated
by the words $a^{k_{1}}b^{l_{1}}...a^{k_{n}}b^{l_{n}}$, $\mathcal{H}%
_{(ab)_{n}}=\mathcal{H}_{a}\otimes \mathcal{H}_{b}\otimes ...\otimes 
\mathcal{H}_{a}\otimes \mathcal{H}_{b}$ and $\mathcal{H}_{(ba)_{n}}$,
defined similarly. Also for any $n$ there exist two invariant $(2n+1)$%
-particle subspaces $\mathcal{H}_{b(ab)^{n}}=\mathcal{H}_{b}\otimes \mathcal{%
H}_{a}\otimes \mathcal{H}_{b}\otimes ...\otimes \mathcal{H}_{a}\otimes 
\mathcal{H}_{b}$ and $\mathcal{H}_{(ab)^{n}a}$, defined similarly. The
spectrum of the hamiltonian reduces using the following equivalence 
\[
\mathcal{H}=\mathcal{H}_{0}\oplus \mathcal{H}_{a}\oplus \mathcal{H}%
_{b}\oplus \mathcal{H}_{ab}\oplus \mathcal{H}_{ba}\oplus ...\oplus \mathcal{H%
}_{(ab)_{n}}\oplus \mathcal{H}_{(ba)_{n}}\oplus \mathcal{H}%
_{b(ab)^{n}}\oplus \mathcal{H}_{(ab)^{n}a}... 
\]
to the spectra of two one-particle hamiltonians, which we have already
studied. For example on $\mathcal{H}_{ab}$ the evolution is given by the
hamiltonian $\mathcal{H}_{a}\otimes 1+1\otimes \mathcal{H}_{b}$.

\subsubsection{Symmetries and quantum grammars}

With some quantum grammars one can associate representations of Lie algebra $%
\mathcal{D}$ of diffeomorphisms of the circle. Denote $D_{n},D_{-n}$ the
hamiltonians, defined by the substitutions $a\rightarrow a^{n+1}$ and \ $%
a^{n+1}\rightarrow a,n\geq 0$, correspondingly. Then in the basis $\nu
_{k}=e_{a^{k}}$ we have for example $D_{n}\nu _{k}=k\nu _{k+n},n\geq 0$.
However for $n<0$ we have only $D_{n}\nu _{k}=(k+n+1)\nu _{k+n}$. For large $%
k$ and fixed $m,n$ one can prove that 
\[
(D_{n}D_{m}-D_{n}D_{m})\nu _{k}=(m-n+o(1))D_{n+m}\nu _{k} 
\]
In this case we say that an asymptotic representation of the algebra $%
\mathcal{D}$ is given. We get more for cyclic words (periodic boundary
conitions). The relation $D_{n}\nu _{k}=k\nu _{k+n}$ takes place for all $%
\left| n\right| <k$, and then 
\[
(D_{n}D_{m}-D_{n}D_{m})\nu _{k}=(m-n)D_{n+m}\nu _{k} 
\]
holds for all $\left| n\right| ,\left| m\right| \ll k$. Otherwise speaking,
the latter equation holds for all $m,n$ everywhere except some finite
dimensional subspace $\mathcal{H}(m,n)$. Note that in both cases the
spectrum of $D_{0}$ coincides with $Z_{+}$.

To get exact representation of the algebra $\mathcal{D}$ one should consider
cyclic words of two symbols $a,a^{-1}$ with relations $aa^{-1}=a^{-1}a=e$,
where $e$ is the empty word, and the substitutions $D_{n}:a\rightarrow
a^{n+1},a^{-1}\rightarrow a^{n+1}$ for all $n\in Z$. Note that one had
always in mind the representation $-V_{0,0}$ in the notation of \cite{katz}.

Consider linear cyclic graphs (that is the words with periodic boundary
conditions) and the grammar on these words with substitutions 
\[
a\rightarrow aa,aa\rightarrow a,aw\rightarrow wa,wa\rightarrow aw 
\]
Select among them the words with one symbol $w$. Note that all cyclic words
of the same length with only one symbol $w$ are equivalent. They generate
the Hilbert space which we denote $\mathcal{H}_{1}$, it is isomorphic to $%
l_{2}(Z_{+})$. Note that in $\mathcal{H}_{1}$\ acts an symptotic
representation of Virasoto algebra.

If one comes from equivalence classes of cyclic words to enumerated words,
that is to the basis $e_{k}\otimes j,j\in Z_{k}$, then hs the action of the
group $Z$ of shifts: on each \ - $e_{k}\otimes j\rightarrow e_{k}\otimes
(j+n),n\in Z,$ where $j+n$\ is taken mod $k$. Thus, the symmetry (earlier
trivial) manifests itself in the enumeration. The action of $Z$ can be
presented with fermionic variables accordingly to the Hooft quantization,
discussed above, see \cite{hooft1, hooft2}.

\end{document}